\documentclass[
reprint,
superscriptaddress,
 amsmath,amssymb,
prd,
]{revtex4-1}

\usepackage{graphicx}
\usepackage{dcolumn}
\usepackage{bm}
\usepackage{amsmath}

\begin{document}

\preprint{APS/123-QED}

\title{A general study of  chameleon fifth force in gravity space experiments}

\author{Martin Pernot-Borr\`as}
\email{martin.pernot\_borras@onera.fr}
\affiliation{DPHY, ONERA, Universit\'e Paris Saclay, F-92322 Ch\^atillon, France}
\affiliation{Institut d'Astrophysique de Paris, CNRS UMR 7095,
Sorbonne Universit\'e, 98 bis Bd Arago, 75014 Paris, France}

\author{Joel Berg\'e}
\affiliation{DPHY, ONERA, Universit\'e Paris Saclay, F-92322 Ch\^atillon, France}

\author{Philippe Brax}
\affiliation{Institut de Physique Th\'eorique, Universit\'e Paris-Saclay, CEA, CNRS, F-91191 Gif-sur-Yvette Cedex, France}

\author{Jean-Philippe Uzan}
\affiliation{Institut d'Astrophysique de Paris, CNRS UMR 7095,
Sorbonne Universit\'e, 98 bis Bd Arago, 75014 Paris, France}
\affiliation{Sorbonne Universit\'e, Institut Lagrange de Paris, 98 bis, Bd Arago, 75014 Paris, France}



\date{\today}

\begin{abstract}
This article investigates the profile of the scalar field  of a scalar-tensor theory subject to  the chameleon mechanism in the context of gravity space missions like the MICROSCOPE experiment. It analyses the experimental situations for  models with an inverse power law potential that can in principle induce a fifth force  inside the satellite, hence either be detected or constrained.  As the mass of the scalar field depends on the local matter density, the screening of the scalar field depends crucially on both the parameters of the theory (potential and non-minimal coupling to matter)  and on the geometry of the satellite. We calculate the profile of the scalar field in 1-, 2- and 3-dimensional satellite configurations without relying on  the thick or thin shell approximations for the scalar field. In particular we consider  the typical geometry with  nested cylinders which is close to the MICROSCOPE design. In this case we evaluate the corresponding fifth force on a test body inside the satellite. This analysis clarifies previous claims on the detectability of the chameleon force by space-borne experiments.



\end{abstract}

\pacs{Valid PACS appear here}
\maketitle

\section{Introduction}

General Relativity (GR) has successfully passed all the experimental tests from the Solar system scales \citep{will_confrontation_2014} to cosmology \cite{Ishak2018}, including the recent confirmation of the existence and properties of gravitational waves \cite{PRLgw2016,abbott_multi-messenger_2017}. However, GR has to be endowed with  a dark sector (including dark matter and a cosmological constant) to provide a cosmological model consistent with observations \cite{Planck2018, PrimCosmo}.  The absence of convincing models for the dark sector  has revived the interest for gravity theories beyond GR \cite{will_confrontation_2014, 2013arXiv1309.5389J}. These theories introduce new degrees of freedom, the effects of which need to be suppressed on small scales although they  may play an important role on cosmological scales.

The simplest extension of GR posits the existence of a non-minimally coupled scalar field. Such a theory, with only one extra degree of freedom, involves at least two free functions (a potential and a universal coupling function when enforcing the weak equivalence principle). These scalar-tensor theories are currently well constrained  from local scale observations \cite{will_confrontation_2014, test_STtheory_gilles} to cosmology \cite{PhysRevD.66.023525,PhysRevD.73.083525}. When the potential and the coupling function enjoy the same minimum these theories can exhibit a cosmological attraction mechanism toward GR in such a way that they are in agreement with local experimental constraints \cite{PhysRevLett.70.2217}. The new degree of freedom can then be considered as a valid dark-energy candidate \cite{PhysRevD.59.123510}.

On small scales, the scalar field is responsible for a fifth force that has to be shielded in order to pass existing experimental tests. Several screening mechanisms have been proposed in the case of scalar-tensor theories, among which the least coupling principle \cite{DAMOUR1994532}, the symmetron \cite{PhysRevLett.104.231301} and the chameleon \cite{khoury_chameleon_2004a,khoury_chameleon_2004} mechanism. The latter model assumes that the coupling and potential functions do not have the same minimum. It follows that the minimum of the effective potential depends on the local density of matter. Hence, in high density environments, the field is heavier and the fifth force may then have a range too small to be detected while in low density environments the fifth force can be long-ranged.

Local gravity experiments on the existence of a fifth force provide already strong constraints on the existence of the chameleon field \cite{burrage_tests_2018,BraxReview}. The main bounds typically come from  atom interferometry \cite{burrageAtom, sabulsky_experiment_2018}, Casimir effect measurements \cite{brax_detecting_2007} or torsion balance experiments to detect short scale forces \cite{upadhye_dark_2012}. Other efforts could lead to new advances  by improving sensitivity or by imagining more original signatures \cite{chiow_multiloop_2018}. It was originally expected \cite{khoury_chameleon_2004a,khoury_chameleon_2004} that space-based experiments could be highly competitive, as they would be performed in a lower density environment.

However, all these experiments  suffer from the problem that their set-ups can screen the fifth force. The recent results on the test of the weak equivalence principle by the MICROSCOPE mission \cite{touboul_microscope_2017} orbiting the Earth have long been expected  to provide new  constraints on chameleon theories (as argued in Refs.~\cite{khoury_chameleon_2004a,khoury_chameleon_2004}). In this experiment, even with a universal coupling, the proof masses can show different screenings of the field leading to different accelerations. As a consequence one expects a violation of the equivalence principle on macroscopic extended objects while it holds at the fundamental level. The question is thus to determine how screened the chameleon field is at the level of a proof mass under the influence of the geometry of a given experiment, a study that has not been performed so far and for which this article is a first step. This is an intricate problem as the distribution of matter is often complex, and given the high non linearity of the chameleon's dynamics. Most of the experiments cited above typically consist of a vacuum cavity enclosed in a shield that can contain experimental devices as electrodes or test masses. As these test masses are extended bodies, they must be taken in account in the profile of the field when computing the force they experience.

Two kinds of effects are expected depending on whether a cavity can be considered isolated or not. On the one hand, in the so-called ``thin-shell'' regime, the field inside the cavity is decoupled from the exterior since the cavity walls exponentially damp the field on a scale smaller than their thickness; in this case, the force applied to a test mass inside the cavity is local and is mostly determined by the structure and the geometry of the cavity. On the other hand, in the so-called ``thick-shell'' regime, the exterior field can penetrate the cavity as it is marginally influenced by the matter constituting the cavity. The limit between those two regimes depends on the model’s parameters and on the geometry of the experiments. In this article, we shall investigate these two dependences and compute the force exerted on a test mass in different settings.

To this end we must determine the chameleon profile inside the experiment. This is a complex problem mostly because of the structure of the boundary conditions and the attraction of the profile toward a fixed point. It has been addressed in various ways in the literature. Analytic models suffer from the non-linearities of the chameleon equation; to overcome them, the Klein-Gordon equation is often approximated by neglecting some terms or by linearizing the chameleon potential \cite{khoury_chameleon_2004, brax_detecting_2007, Brax:2013cfa, PhysRevD.87.105013,  burrage_probing_2015, burrage_proposed_2016, ivanov_exact_2016, nakamura_chameleon_2018, kraiselburd2018, kraiselburd2019}. Numerical models \cite{upadhye_dark_2012, hamilton_atom-interferometry_2015, elder_chameleon_2016,schlogel_probing_2016, burrage_shape_2018} suffer from the limited resources they have, leading to solving the equation in a bounded region, setting the boundary conditions at a finite distance or  neglecting some short-scale variations of the field. Besides the fact that this last point may lead to an incorrect field even where the field varies slowly, this is very problematic for experiments using extended test masses. Short scale variations are indeed more likely to happen in matter, impacting on  the very gradient responsible for the force  aimed to be measured. This caveat is also encountered in analytic approaches.

This article overcomes these approximations. We tackle the problem numerically and consider all terms of the chameleon equation. To comply with the necessity to set boundary conditions at infinity, we consider a low-density background  environment in which we embed a high-density system whose complexity increases throughout the paper. Our final goal is to approach the concentric-cylinder geometry of the MICROSCOPE instrument \cite{touboul_microscope_2017}.
Although we restrict ourselves to static configurations with symmetries simpler than in realistic cases,  this paper will  pave the way to further study  including asymmetries and dynamics. We should note that most configurations studied in this article have already been partly explored in the literature, whether in specific regimes or with assumed boundary conditions. Here we investigate general profiles to clarify the boundary condition problem and to infer robust criteria to legitimize the approximations encountered in the literature.

This article is organized as follows. The first part of the paper focuses on one-dimensional geometries. In Section \ref{sec:init}, we discuss the dynamics of the chameleon field paying particular attention to the role of boundary conditions. In Section \ref{sec:wallsec} we analyze the case of an infinite wall and in Section \ref{sec:case2} we consider the case of a one dimensional cavity. Following these 1D configurations, we explore 2D and 3D symmetrical configurations in Section \ref{sec:2D3D}. Finally in Section \ref{sec:appli}, we notice that the exact numerical integration of the field profile in a cavity leads to discrepancies with the analytic approximations used to evaluate the  Casimir pressure induced by the chameleon field. We also consider the effect of the chameleon force on the motion of atoms in a cavity and the corresponding drifting time which could serve  as a testing ground for such models. Finally we present the field profile in nested cylindrical configurations close to the MICROSCOPE setting as a first step toward a more thorough investigation of the constraints from MICROSCOPE on chameleons which are left for future work. We conclude in Section \ref{sec:concl}.

\section{The chameleon's profile and initial conditions}
\label{sec:init}
\subsection{Theoretical model}
The chameleon mechanism is given in the Einstein frame by
\begin{equation}
\begin{split}
S = \int {\rm d}x^4 ~ &\sqrt{-g} \left[\frac{{\rm M}_{ \rm Pl}^2}{2}{\displaystyle R} -  \frac{1}{2}\partial^\mu\partial_\mu\phi - V(\phi)\right]\\
&- \int {\rm d^4 }x \mathcal{L}_{\rm m}(\widetilde{g}_{\mu\nu}, \phi, ...),
\end{split}
\end{equation}
where $\phi$ is the chameleon field, $V$ its potential, ${\rm M}_{ \rm Pl}$ the reduced Planck mass, ${\displaystyle R}$ the Ricci scalar, $g_{\mu\nu}$ the Einstein frame metric, $g$ its determinant and $\mathcal{L}_{\rm m}$ the matter Lagrangian. The field couples non-minimally to matter through the Jordan frame $\widetilde{g}_{\mu\nu} = A^2(\phi)g_{\mu\nu}$, where $A$ is a universal coupling function. We define the dimensionless coupling constant $\beta = {\rm M}_{ \rm Pl} \frac{{\rm dln}A}{{\rm d}\phi}$. The field could have different coupling function for each component of matter, but here we restrict to a universal coupling.

For static configurations of non relativistic matter, the field follows the Klein-Gordon equation
\begin{equation}
\nabla^2 \phi = V_{\rm eff,\phi} \equiv V_{,\phi} + \frac{\beta}{{\rm M}_{\rm Pl}} \rho_{\rm mat},
\label{KGeq}
\end{equation}
where $\rho_{\rm mat}$ is the mass density function. For non-static configurations, the Laplacian would be a d'Alembertian. We use the Ratra-Peebles inverse power-law potential of energy scale $\Lambda$ and exponent $n$ \cite{ratra_cosmological_1988, burrage_tests_2018} as a typical example of chameleon model
\begin{equation}
V(\phi) = \Lambda^4\left(1+\frac{\Lambda^{n}}{\phi^n}\right).
\end{equation}

The effective potential $V_{\rm eff}$ has a minimum given by
\begin{equation}
\phi_{\rm min} (\rho_{\rm mat}) = \left( {\rm M}_{ \rm Pl} \frac{n \Lambda^{n+4}}{\beta \rho_{\rm mat}} \right)^{\frac{1}{n+1}}.
\label{phimin}
\end{equation}
It plays a central role in the chameleon dynamics.

We recall that in a medium with constant density, the field is expected to relax exponentially to the minimum of its potential. It varies on a typical scale of the order of its local Compton wavelength
\begin{equation}
\lambda_{\rm c}(\rho_{\rm mat}) \equiv m^{-1}(\rho_{\rm mat}) = \frac{1}{\sqrt{V''_{\rm eff}(\phi_{\rm 	min})}},
\end{equation}
which is explicitly given, in the models considered in this article by
\begin{equation}
\lambda_{\rm c}(\rho_{\rm mat}) = \sqrt{\frac{1}{n (n + 1) \, \Lambda^{n+4}} \left(\frac{n \,{\rm M}_{\rm Pl} \, \Lambda^{n+4}}{\beta \, \rho_{\rm mat}}\right)^\frac{n+2}{n+1}}.
\end{equation}

The fifth force induced by the coupling to the chameleon field on a point test mass is proportional to the gradient of the scalar field and given by
\begin{equation}
\vec{F} = - \frac{\beta}{{\rm M}_{ \rm Pl}} m_{\rm test} \vec{\nabla}\phi.
\label{forceq}
\end{equation}
Nevertheless an extended body cannot, a priori, be considered as a test body since its own matter density  impacts the field profile inside its volume. Hence, to evaluate properly the force, one needs to solve consistently for the field profile including the extended body and to integrate this force over the whole volume of the body.

In what follows, it is convenient to re-write the  chameleon's Klein-Gordon equation \eqref{KGeq} in terms of $\phi_{\rm min}$ as
\begin{equation}
\nabla^2 \phi = n \Lambda^{n+4} \left[ \frac{1}{\phi_{\rm min}^{n+1}(\rho_{\rm mat})}-\frac{1}{\phi^{n+1}}\right],
\label{ChamEq}
\end{equation}
where the dependence to the local mass density is now contained in $\phi_{\rm min}$.

If we consider a region of space with local density $\rho_{\rm vac}$, large compared to the corresponding chameleon's Compton wavelength and far from any perturbing body, we can assume that the field is uniform with a value $\phi_{\rm vac} = \phi_{\rm min}(\rho_{\rm vac})$ . We shall now study the way a 1-dimensional material structure affects this uniform profile as experiencing different $\phi_{\rm min}$ the field should depart from $\phi_{\rm vac}$.

\subsection{Initial conditions in one dimension}
\label{sec:chamdyn}
The chameleon profile is solution to a boundary value problem. Given the previous discussion, the field shall relax to its minimum value in the external space, such that
\begin{equation}
\begin{cases}
\phi \xrightarrow[x \to +/- \infty]{} \phi_{\rm min}(\rho_{\rm vac})\\
\phi'  \xrightarrow[x \to +/- \infty]{} 0\\
\end{cases}.
\label{boundarycond}
\end{equation}

Such a boundary value problem can be solved using finite difference methods. However, due to the finite extent of computational memory we cannot set boundary conditions at infinity. We would need then to set the boundary conditions at a finite distance of the considered object, and make a compromise between computational memory limits and the distance at which we can consider that the gap between the value the field takes and $\phi_{\rm vac}$ becomes negligible. The Compton wavelength in  vacuum $\lambda_{\rm c,vac}$ is an estimate of this distance \cite{khoury_chameleon_2004, burrage_tests_2018}. This is an approximate criterion, a more accurate one will be determined in the following by  direct integration as an initial value problem.

\vspace{0.5cm}
Initial conditions cannot be chosen to be at $\phi_{\rm vac}$ with a null derivative. To understand this  we must note the key role played by $\phi_{\rm min}$, as a fixed point of the theory. One can check that for $n>0$, we have
\begin{equation}
\frac{{\rm d}^2\phi}{{\rm d}x^2} \,
\begin{cases}
       > 0 &\quad\text{if} ~\phi > \phi_{\rm min}\\
       = 0 &\quad\text{if} ~\phi = \phi_{\rm min}\\
       < 0 &\quad\text{if} ~\phi < \phi_{\rm min}\\
\end{cases},
\label{dyncond}
\end{equation}
so that the field derivative is increasing (resp. decreasing) for $\phi > \phi_{\rm min}$ (resp. $\phi < \phi_{\rm min}$). For $\phi = \phi_{\rm min}$, the field's derivative will not vary.

\begin{figure}
\includegraphics[width = 0.70\columnwidth]{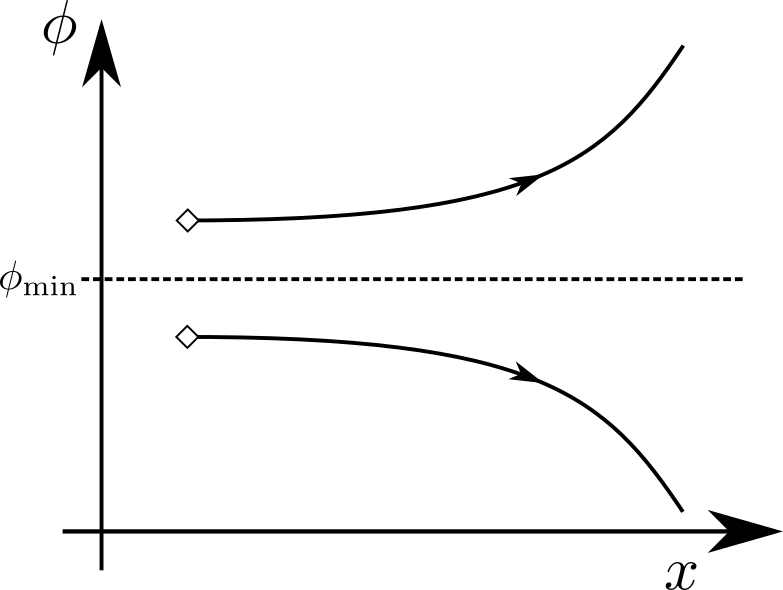}
\caption{Sketch of the field profiles for null initial derivatives: $\phi_{\rm i}' = 0$. Different behaviors are obtained depending on the magnitude of $\phi_{\rm i}$ compared to $\phi_{\rm min}$.}
\label{ChamDynI}
\end{figure}

Hence if we choose as initial conditions $\phi_{\rm i}' = 0$, as in Fig.~\ref{ChamDynI}, the field will diverge monotonically toward $+\infty$ or $-\infty$ at large $x$, for an initial value $\phi_{\rm i} > \phi_{\rm min}$ or $\phi_{\rm i} < \phi_{\rm min}$ respectively. For an initial value $\phi_{\rm i} = \phi_{\rm min}$, the field being at the fixed point remains constant.

\begin{figure}
\includegraphics[width = 0.8\columnwidth]{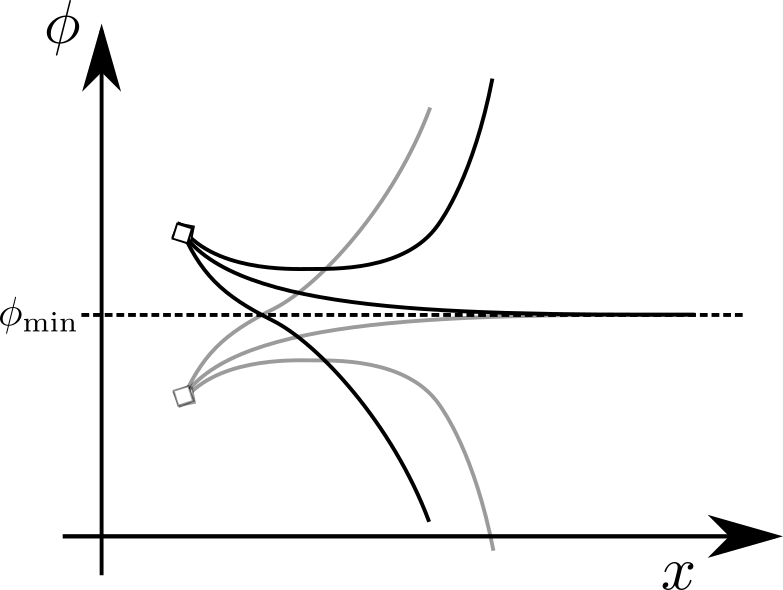}
\caption{Sketch of the field profiles for non-zero initial derivative: $\phi_{\rm i}' \neq 0$. Different behaviors -- each line -- are obtained depending on the magnitude and sign of $\phi_{\rm i}'$. Grey lines correspond to $\phi_{\rm i} < \phi_{\rm min}$.}
\label{ChamDynII}
\end{figure}

If we choose $\phi_{\rm i}' \neq 0$, considerations in Eq.~\eqref{dyncond} do not change, the $\phi'$ evolution remains the same. Nevertheless the field evolution will no longer be monotonic, and eventually show maxima and minima. In the case where  $\phi_{\rm i} > \phi_{\rm min}$, if $\phi_{\rm i}' > 0$, the field will diverge more rapidly than if $\phi_{\rm i}' = 0$.  The different possible evolutions for $\phi_{\rm i}' \neq 0$ are sketched in Fig.~\ref{ChamDynII}.  For small values of $|\phi_{\rm i}'|$, the field has not enough ``speed'' to reach $\phi_{\rm min}$, thus it will reach a minimum and then diverge. For high values of $|\phi_{\rm i}'|$, the field can reach $\phi_{\rm min}$. When crossing $\phi_{\rm min}$, $\phi_{\rm}'$ will still be negative, but as we would have now $\phi < \phi_{\rm min}$, it will decrease, and make the field diverge negatively.
For a given $\phi_{\rm i}$, there is one and only value of $\phi_{\rm i}'$, in between these two behaviors, that will lead $\phi'$ to vanish precisely when the field reaches $\phi_{\rm min}$. In this case, $\phi_{\rm i}$ is fixed by the considered matter distribution.

 Note that the case where  $\phi_{\rm i} < \phi_{\rm min}$ is completely symmetric to the case where $\phi_{\rm i} > \phi_{\rm min}$ as the light grey curves in Fig.~\ref{ChamDynII} show.

\vspace{0.5cm}
In 1D, the problem can be treated relatively easily. The chameleon equation  can indeed be integrated once, from infinity -- where boundary conditions are verified -- to the place we want to set the initial conditions. This gives a condition on $\phi_{\rm i}'$ in terms of $\phi_{\rm i}$

\begin{equation}
\frac{1}{2} \phi^{'2}_{\rm i} = \frac{n}{\phi_{\rm vac}^{n+1}}(\phi_{\rm i}-\phi_{\rm vac}) + \left( \frac{1}{\phi_{\rm i}^n} - \frac{1}{\phi_{\rm vac}^n} \right).
\label{primecond}
\end{equation}

This leaves us with only one initial parameter to deal with. We can use shooting methods, varying $\phi_{\rm i}$ to obtain the proper solution for the considered configuration.

\section{Effect of an infinite wall on the chameleon's dynamics}
\label{sec:wallsec}
\subsection{An interface between two infinite domains}
\label{sec:case0}
As a first step, we consider the simple case of an interface between two infinitely extended domains of different densities, for instance a high-density wall and a low-density vacuum of density  $\rho_{\rm wall}$ and $\rho_{\rm vac}$, respectively.

Far from the interface, the field will tend toward the value that minimizes the potential in each environment, $\phi_{\rm wall}$ and $\phi_{\rm vac}$ respectively. Note that Eq.~\eqref{phimin} implies $\phi_{\rm vac} > \phi_{\rm wall}$. In between, the field will evolve smoothly and cross the interface with a value $\phi_{\rm I}$ and a continuous derivative, with $\phi_{\rm wall} < \phi_{\rm I} < \phi_{\rm vac}$. To solve for the profile numerically, we set the initial conditions at this interface. In the wall as here $\phi_{\rm I} > \phi_{\rm wall}$, we are in the case shown by the black line that asymptotically tends toward $\phi_{\rm min}$ in Fig.~\ref{ChamDynII}. In the other domain, the symmetric dotted line is more relevant, as now $\phi_{\rm I} < \phi_{\rm vac}$.

In this configuration, no shooting methods are required. This is because the asymptotic conditions on both sides of the interface give two different conditions, equivalent to Eq.~\eqref{primecond}, on $\phi_{\rm I}$ and $\phi_{\rm I}'$ given by
\begin{equation}
\frac{1}{2} \phi^{'2}_{\rm I} = \frac{n}{\phi_{\rm vac}^{n+1}}(\phi_{\rm I}-\phi_{\rm vac}) + \left( \frac{1}{\phi_{\rm I}^n} - \frac{1}{\phi_{\rm vac}^n} \right) \, \, \, \, \, \, \,
\end{equation}
\vspace{-0.5cm}
\begin{equation}
\frac{1}{2} \phi^{'2}_{\rm I} = \frac{n}{\phi_{\rm wall}^{n+1}}(\phi_{\rm I}-\phi_{\rm wall}) + \left( \frac{1}{\phi_{\rm I}^n} - \frac{1}{\phi_{\rm wall}^n} \right) \, \, \, \, .
\end{equation}

\begin{figure}
\includegraphics[width = 0.9\columnwidth]{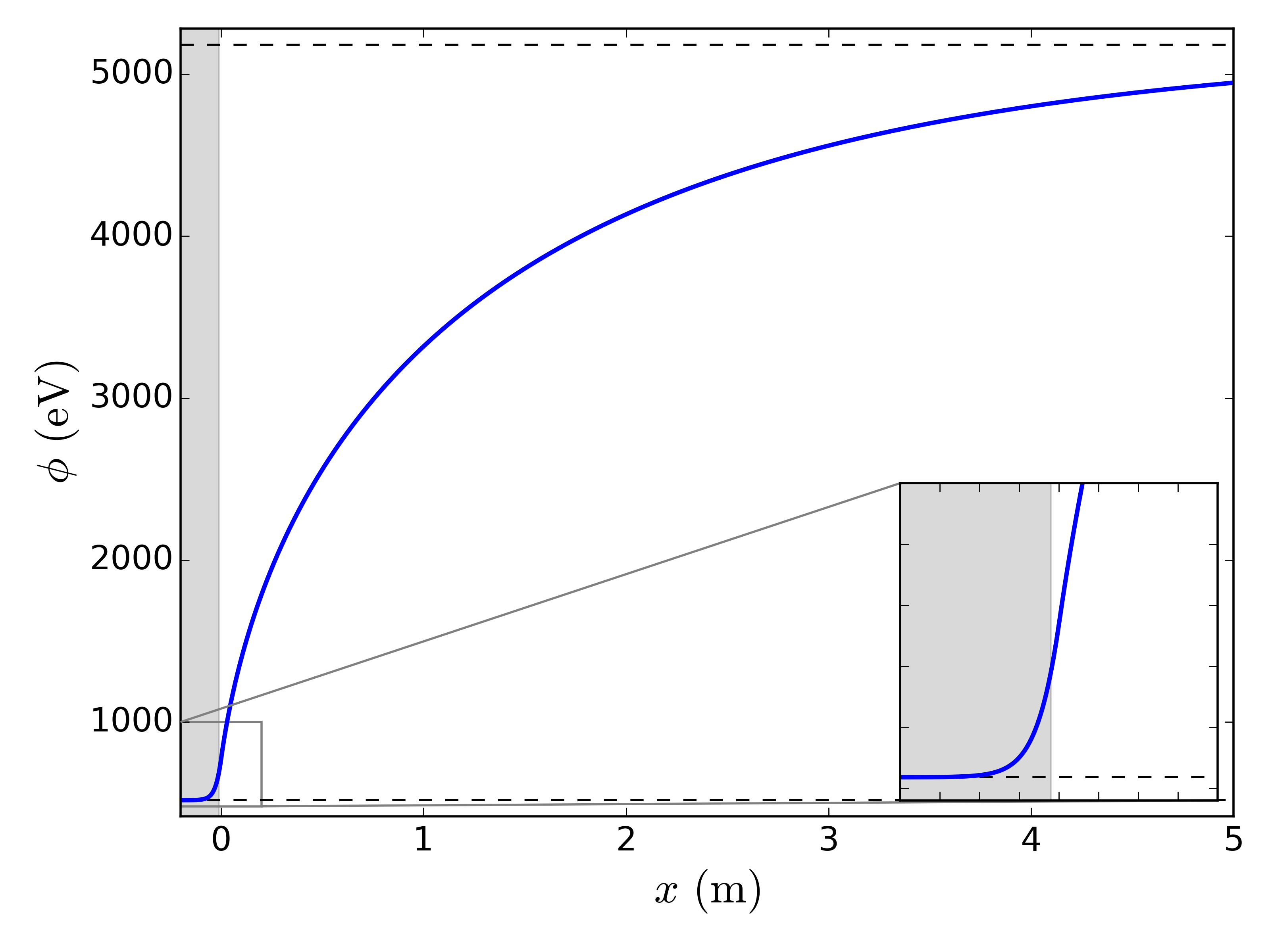}
\caption{Example of field profiles with interface at $x=0$. $\phi_{\rm min}$ values are shown with the two dashed lines. The grey zone is the higher density domain.}
\label{profcase0}
\end{figure}

Combining these two equations gives $\phi_{\rm I}$ and $\phi'_{\rm I}$ in terms of $\phi_{\rm wall}$ and $\phi_{\rm vac}$. We can then integrate numerically in both domains. Figure \ref{profcase0} depicts such a solution with the interface at $x=0$. Note that for this profile and for every other profile computed in the following, if not stated otherwise, we consider the case where: $n=2$, $\beta = 1$, $\Lambda = 1 \, {\rm eV}$, $\rho_{\rm wall} = 8.125 \,{\rm g.cm}^{-3}$, $\rho_{\rm vac} = 10^{-3} \rho_{\rm wall}$ ($\phi_{\rm vac} = 10 \,\phi_{\rm wall}$, for $n=2$).
In each domains, the field reaches the corresponding minimum of its potential within scales given by the Compton wavelength $\lambda_{\rm c}(\rho_{\rm mat})$. For the set of parameters and densities considered throughout the article, we have $\lambda_{\rm c, vac} \simeq 2 \, {\rm m}$ and $\lambda_{\rm c, wall} \simeq 0.02 \, {\rm m}$. Note that for the sake of clarity, we chose $\rho_{\rm vac}$ and $\rho_{\rm wall}$ not to be vastly different. For more realistic vacuum cavities, $\rho_{\rm vac} = 10^{-15} \rho_{\rm wall}$, implying a more significant difference between $\lambda_{\rm c, vac}$ and $\lambda_{\rm c, wall}$.

\subsection{A single wall}
\label{sec:case1}

We then consider a single wall of uniform density embedded in the low-density background  environment. We denote by $e$ its thickness. On both sides of the wall, the field will evolve similarly as the previous section. We set the initial conditions on one of the borders of the wall. Say we choose to set them of the right side. We denote them by $\phi_{\rm e}$ and $\phi'_{\rm e}$. By symmetry, the field value will be the same on  the other border of the wall, with a derivative of opposite sign. As in the previous section, we know $\phi_{\rm wall} < \phi_{\rm e} < \phi_{\rm vac}$, with $\phi_{\rm e}' > 0$ and the boundary conditions gives by  direct integration a condition on $\phi^{'}_{\rm e}$ in terms of  $\phi_{\rm e}$,
\begin{equation}
\phi^{'}_{\rm e} = \pm \sqrt{2 \left[\frac{n}{\phi_{\rm vac}^{n+1}}(\phi_{\rm e}-\phi_{\rm vac}) + \left( \frac{1}{\phi_{\rm e}^n} - \frac{1}{\phi_{\rm vac}^n} \right)\right]},
\end{equation}
where we choose the positive sign in this case.

If we look toward the wall, the initial field derivative $\phi_{\rm e}'$ will look negative. As $\phi_{\rm e}
> \phi_{\rm wall}$, the field will be similar to the case that shows a minimum in Fig.~\ref{ChamDynII}. The field will then evolve from $\phi_{\rm e}$ to a minimum value reached at the center of the wall. The scale of this evolution will depend on the magnitude of $\phi_{\rm e}$. Consequently there is a one-to-one mapping between $e$ and $\phi_{\rm e}$. The larger $\phi_{\rm e}$, the smaller $e$.

\begin{figure}
\includegraphics[width = 0.9\columnwidth]{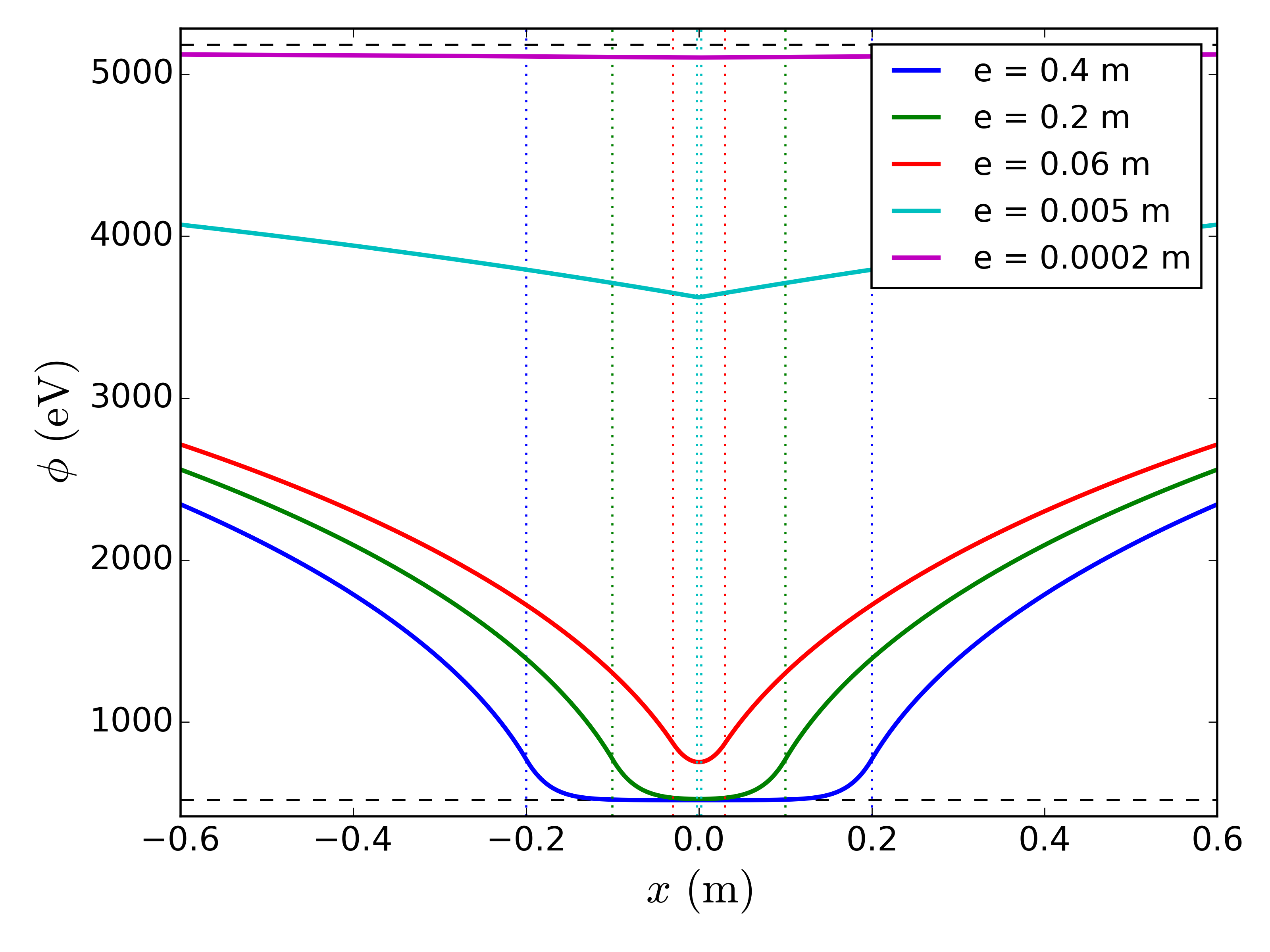}
\caption{Field profiles for different wall thickness $e$. The two value of $\phi_{\rm min}$ are shown with the two dashed lines. Dotted lines show the extent of walls.}
\label{profcase1}
\end{figure}

Figure \ref{profcase1} depicts the numerical integration of a series of profiles  for different $e$. Dotted lines delimit the frontiers of the considered wall. As expected the thicker the wall gets, the more space the field has to evolve inside the wall, so the closer it gets to $ \phi_{\rm wall}$.

\subsubsection{${\phi_{\rm e} (e)}$ relation}
\begin{figure}
\includegraphics[width = 0.9\columnwidth]{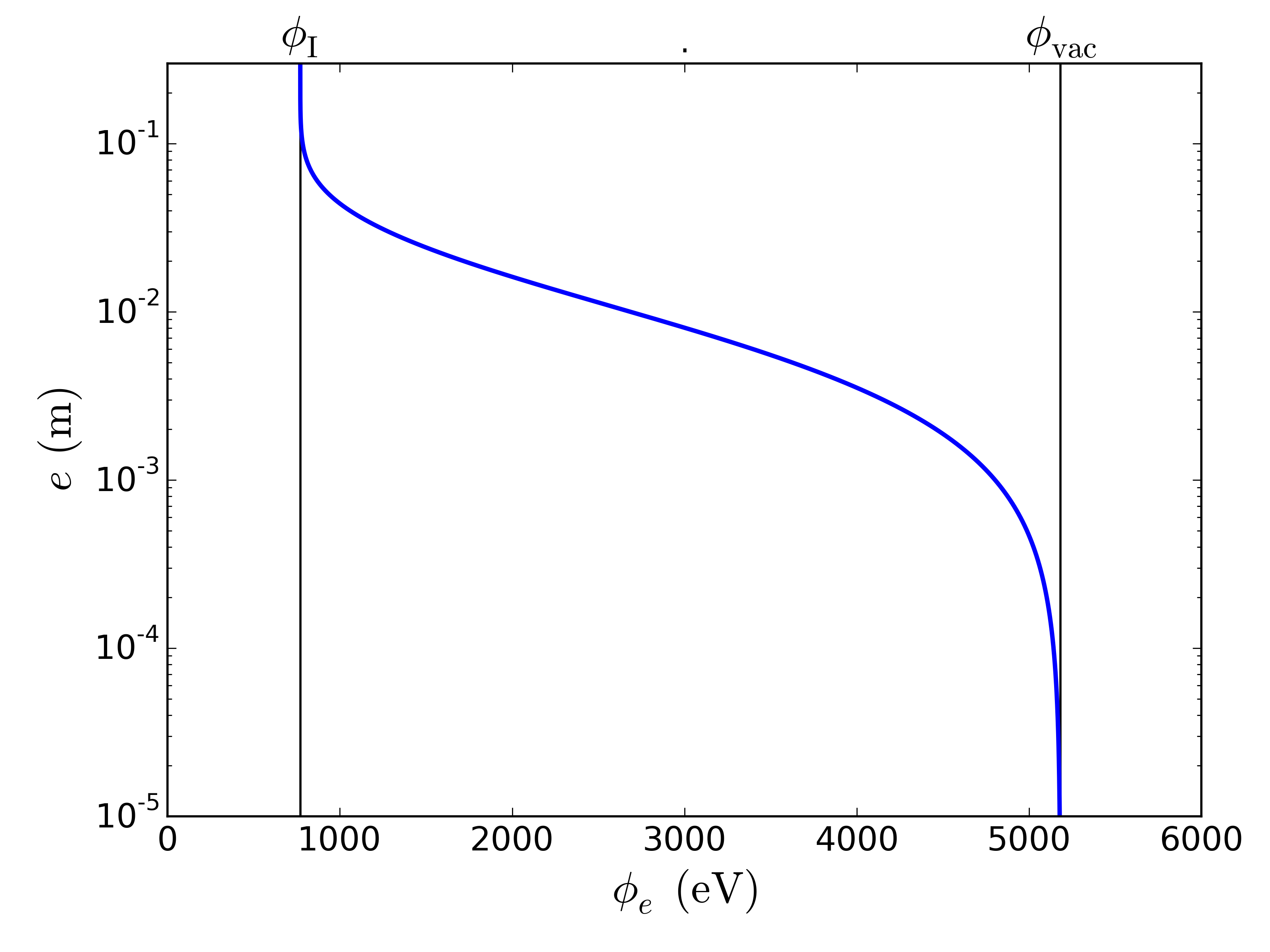}
\caption{Example of relation $\phi_{\rm e} (e)$. The black lines denote $\phi_{\rm I}$ and $\phi_{\rm vac}$.}
\label{exfigrel}
\end{figure}

To compute the profile associated to any wall thickness, we need to determine the relation $\phi_{\rm e} ({e})$ that can be obtained by a shooting method. Figure \ref{exfigrel} shows an example a such a relation for our fiducial parameters ($n, \beta, \Lambda$), and $\phi_{\rm wall}$, $\phi_{\rm vac}$.

This figure shows that all possible values for $e \in \mathbb{R}^+$ are spanned by a limited range for $\phi_e \in \left[ \phi_{\rm I}, \phi_{\rm vac} \right]$ given by two limiting regimes:

- $\phi_{\rm vac}$ corresponds to the limiting case where the wall becomes infinitely thin and represents a very tiny perturbation to the background field.

- $\phi_{\rm I}$ corresponds to the other limit case where the field tends to reach $\phi_{\rm wall}$ at the center of the wall: we say the field is completely screened inside the wall. The profile can be seen as two concatenated profiles of the Section \ref{sec:case0} case, which explains the value $\phi_{\rm I}$ as lower boundary. This behavior is consistent with the fact that the field is exponentially suppressed in the wall on scales of Compton wavelength $\lambda_{\rm c,wall}$ in the wall.

\subsubsection{${\phi_{\rm e} (e)}$'s dependence on $\Lambda$ and $\beta$\\}

\begin{figure}
\includegraphics[width = 0.9\columnwidth]{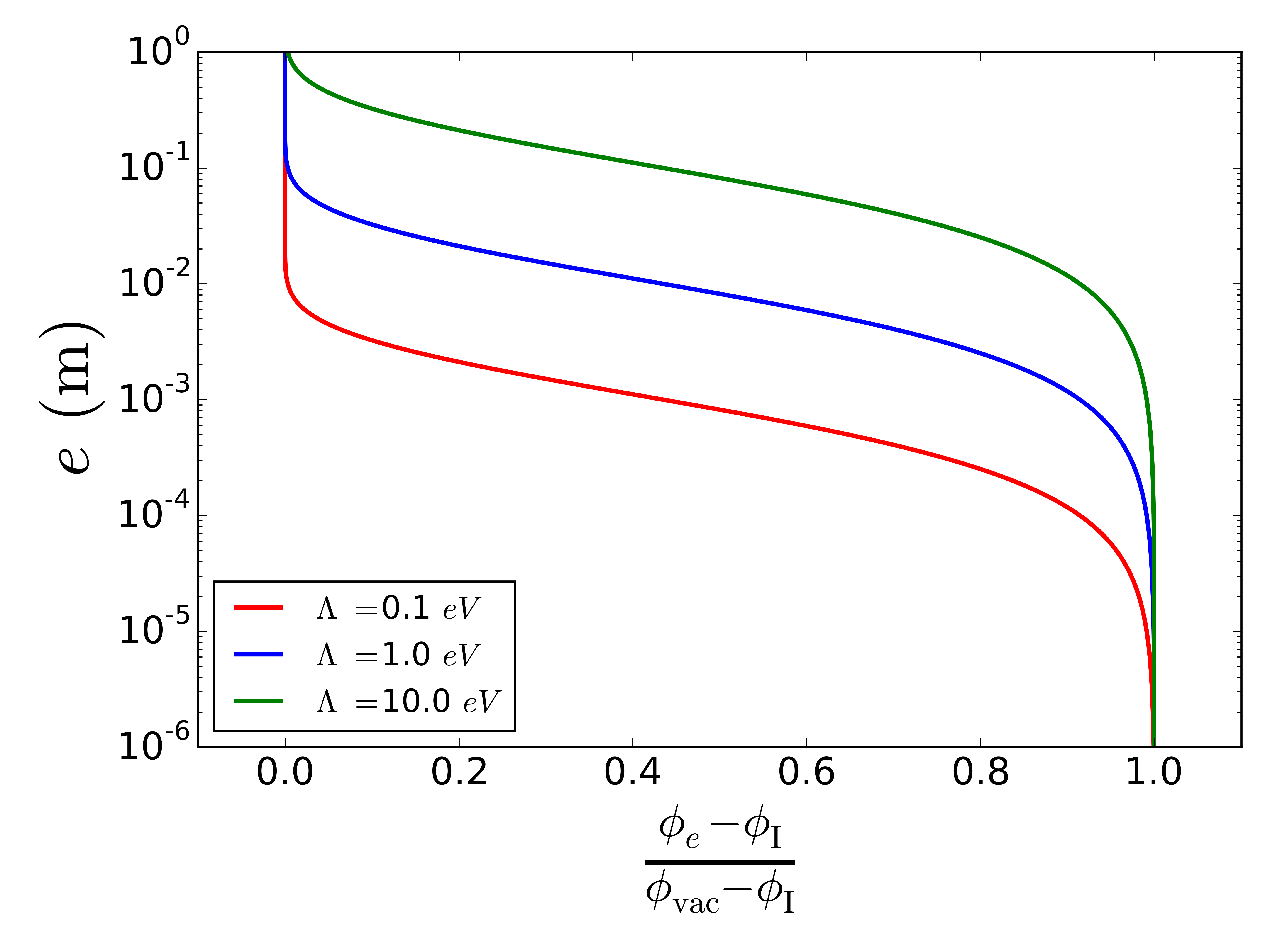}
\caption{Variation of the $\phi_{\rm e} (e)$ with $\Lambda$, for  $\Lambda = 0.1\,, 1\,, 10\,{\rm eV}$. The interval on which $\phi_{\rm e} (e)$ is defined is normalized to [0,1].}
\label{evarlbda}
\end{figure}

$\phi_{\rm I}$ and $\phi_{\rm vac}$ depend on $\Lambda$ and $\beta$ in such a way that the interval $\left[\phi_{\rm I}, \phi_{\rm vac} \right]$ spreads or shrinks. It spreads logarithmically with $\Lambda$ and shrinks logarithmically with $\beta$.  Figure \ref{evarlbda} shows how the $\phi_{\rm e} (e)$ relation depends on $\Lambda$. Here the interval $\left[ \phi_{\rm I}, \phi_{\rm vac} \right]$ is normalized to an interval $\left[ 0,1 \right]$.

This figure shows that when varying $\Lambda$, the $\phi_{\rm e} (e)$ relations have the same slope, but are just shifted on the $e$-axis. The dependence on $\beta$ is similar, albeit in the opposite direction. To understand this variation, we can choose a specific value in the  $\left[\phi_{\rm I}, \phi_{\rm vac} \right]$ interval, say $\frac{\phi_{\rm e}-\phi_{\rm I}}{\phi_{\rm vac}-\phi_{\rm I}} = 0.5$, and see how $e$ varies with $\Lambda$ and $\beta$. We can fit this variations as
\begin{equation}
e (\Lambda, \, \beta) = A \times \Lambda \times \beta^{-\frac{2}{3}},
\end{equation}
where $A$ is a coefficient that depends in a non trivial way on $\rho_{\rm wall}$ and $\rho_{\rm vac}$. In the cases considered in this figure, $\rm{A} = 2.15 \times 10^{-3} \,\rm{m.eV^{-1}}$.

\subsubsection{Screening of the wall}
\label{sec:screening}
As mentioned before, when the wall gets thicker, it gets screened so that the field tends to the value that minimizes the potential inside the wall $\phi_{\rm wall}$ at the center of the wall. In this case, we can consider that the field's dynamics on both side of the wall decouple, such that if the matter distribution were to change on one side of the wall, this would not influence the field on the other side. This will be important for the case of a cavity.

\begin{figure}
\includegraphics[width = 0.85\columnwidth]{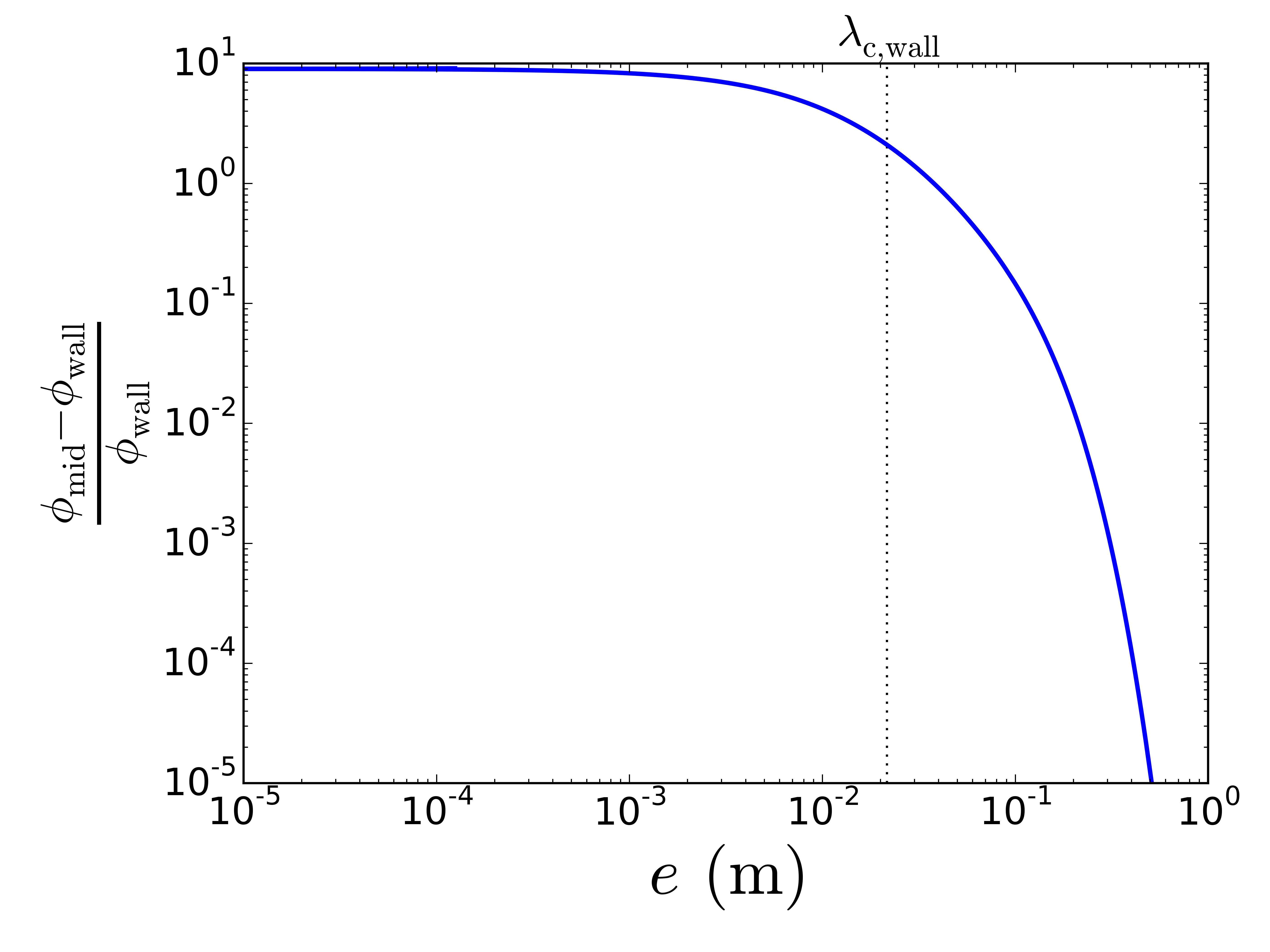}
\caption{Variation with the wall thickness $e$ of the difference between the value of the field at the center of the wall and $\phi_{\rm wall}$.}
\label{phimid}
\end{figure}

This was expected to happen for walls thicker than $\lambda_{\rm c,wall}$ \cite{burrage_tests_2018}. Nevertheless, we can deduce from our simulations a more accurate criterion. We can indeed measure the difference between $\phi_{\rm wall}$ and the effective minimum value the field reaches at the center of the wall. Figure \ref{phimid} shows its evolution with the wall thickness.

As expected we observe this difference slowly decreases when the wall gets thicker. It then suddenly drops down when the wall thickness exceeds $\lambda_{\rm c,wall}$. We can consider this gap gets negligible when it reaches a thickness of roughly $100 \, \lambda_{\rm c,wall}$, as it gets smaller than typical numerical precisions. This criterion is useful for other numerical methods such as finite difference methods, in which one can only solve the field in a bounded region. For instance, when considering a system totally surrounded by walls, one could safely set initial conditions for the field to be at its minimum deeply inside these walls, as long as these wall have thickness greater than $100 \, \lambda_{\rm c,wall}$.

\subsubsection{Range of influence of a wall}

\begin{figure}
\includegraphics[width = 0.80 \columnwidth]{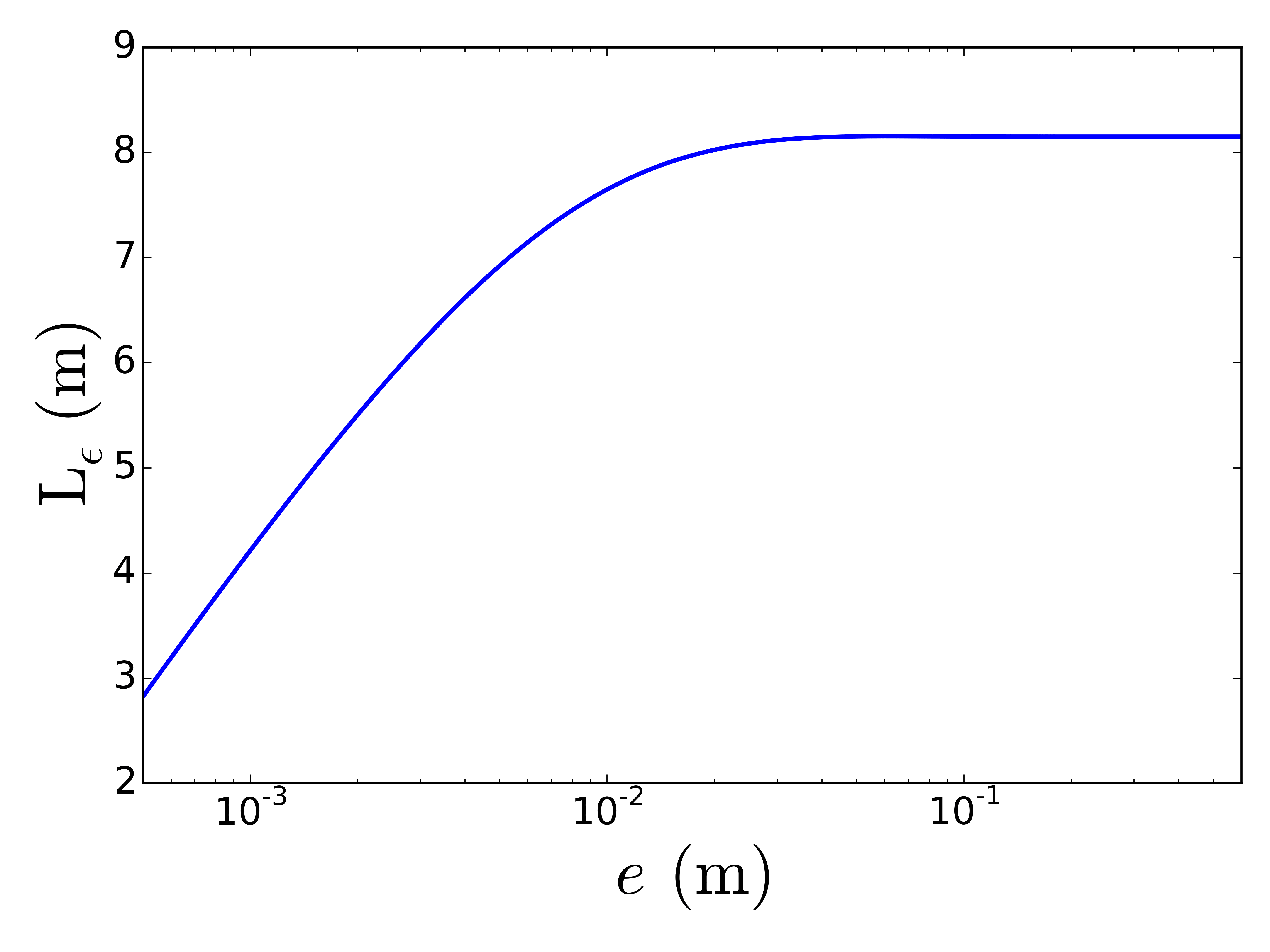}
\caption{Scale of influence $L_\epsilon$ of a wall as a function of $e$, for $\epsilon = 1\%$. }
\label{LtoNegl}
\end{figure}

We can also deduce a scale of influence of a wall. Outside the wall, the field slowly relaxes to its asymptotic value $\phi_{\rm vac}$.

The typical  relaxation scale $L_\epsilon$ at which the gap between the field and $\phi_{\rm vac}$ gets negligible, solves
\begin{equation}
\frac{\phi(e/2 + L_\epsilon)-\phi_{\rm vac}}{\phi_{\rm vac}} = \epsilon,
\end{equation}
where we take $\epsilon$ to be small. We can then consider that for distances to the wall larger than $L_\epsilon$, the dynamics of the field are no longer influenced by the wall.

Figure \ref{LtoNegl} shows how this scale of influence varies with the wall thickness, for $\epsilon = 1\%$. We observe it increases when the wall gets thicker, and finally reaches a plateau when the wall starts being totally screened i.e. its thickness exceeds $\lambda_{\rm c,wall}$.

We can infer a useful criterion from such a figure. We can safely consider that we are out of the influence of the wall, when distant by more than $10 \, \lambda_{\rm c,vac}$ from it. 	

\section{One dimensional cavity}
\label{sec:case2}

\begin{figure}[h!]
\includegraphics[width = 0.65\columnwidth]{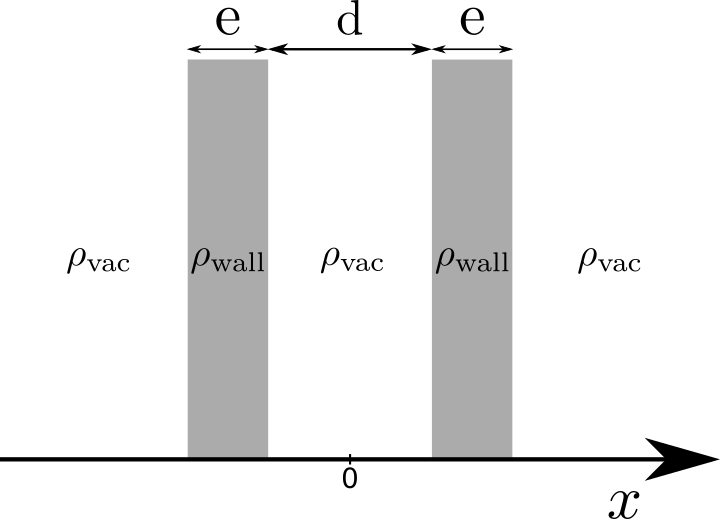}
\caption{1D cavity.}
\label{schcase2}
\end{figure}

\subsection{Profile and $\bm{\phi_{\rm d}(d)}$ relation}

The experimental case of a cavity in one dimension is modeled as two walls of equal thickness $e$, separated by an empty space of size $d$, as illustrated in Fig.~\ref{schcase2}. For simplicity we assume the cavity to have the same density as the background environment $\rho_{\rm vac}$.

We follow the same approach as in the previous section. We impose initial conditions at the external border of a wall, say the right one. When fixing the thickness of the wall, the initial conditions will be determined by the size of the cavity $d$, so we denote them by $\phi_{\rm d}$ and $\phi'_{\rm d}$. The same first integration in the external vacuum region gives a condition on  $\phi'_{\rm d}$ in terms $\phi_{\rm d}$, to satisfy boundary condition at infinity. The magnitude of $\phi_{\rm d}$ determines the dynamics of the field inside the walls/cavity system. The overall profile will still be symmetric around the cavity center.

Inside the walls, the field is no longer symmetric. It must indeed reach a value smaller than $\phi_{\rm d}$ on the inside border of the wall, as otherwise it would have the same asymptotic behavior as in the external vacuum region or diverge. Thus if $\phi_{\rm e}(e)$ is the initial value of field given in the previous section for a wall of thickness $e$, we should now choose $\phi_{\rm d} < \phi_{\rm e}(e)$. This way the field will not have enough  ``speed" to reach $\phi_{\rm d}$ again at the border of the cavity, but it will instead reach a value $\phi(d/2) < \phi_{\rm d} < \phi_{\rm vac}$, with a positive derivative. Then in the cavity the field will have the same kind of dynamics with a maximum as in the bottom grey line in Fig.~\ref{ChamDynII}, and reach $\phi(d/2)$ again at the other side of the cavity.

For a fixed wall thickness $e$, $\phi_{\rm d}$ will determine the value $\phi(d/2)$, which will determine  the maximum field value  $\phi_0$ at  the cavity center. Thus we will obtain the corresponding cavity size $d$. We use the same shooting method as in the previous section to determine numerically how $\phi_{\rm d}$ varies with $d$, and hence the $\phi_{\rm d}(d)$ relation. Note that the larger d, the larger $\phi_0$.

\begin{figure}
\includegraphics[width = 1\columnwidth]{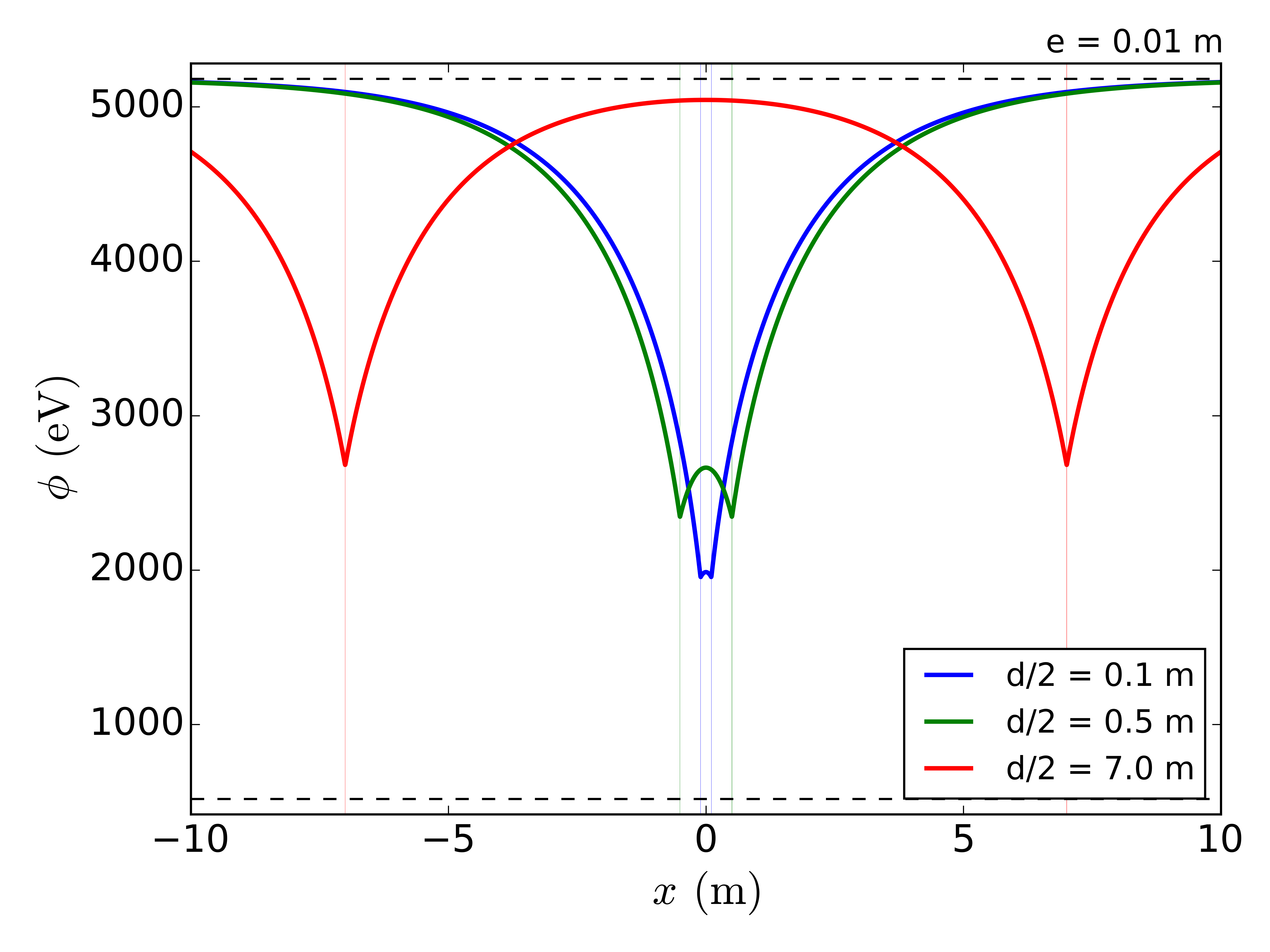}
\includegraphics[width = 1\columnwidth]{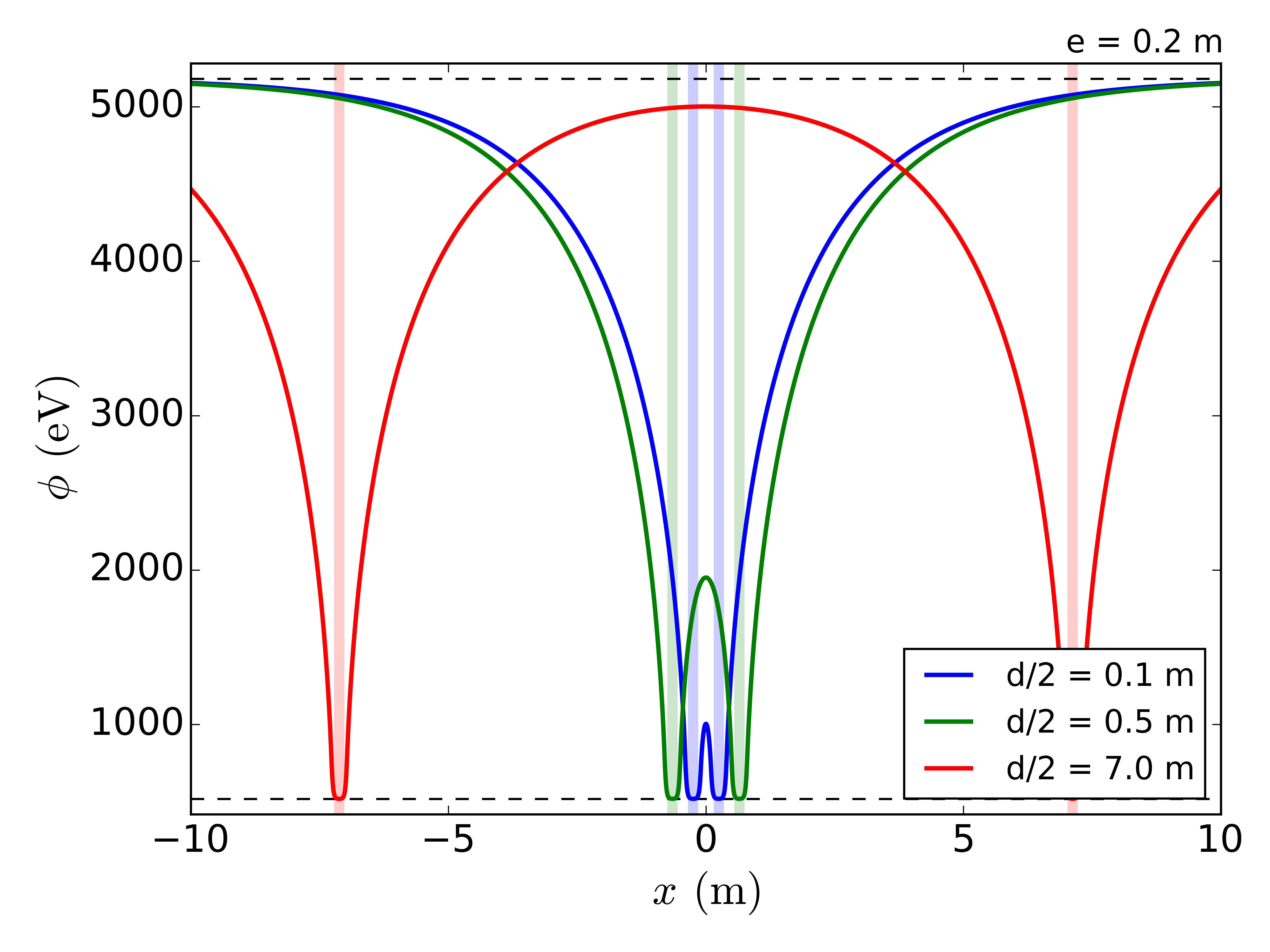}
\caption{Field profiles for different cavity sizes with un-screened (upper panel) and screened (lower panel) walls of thickness $e = 0.01\,{\rm m}$ and $e = 0.2{\rm m}$ respectively. $\phi_{\rm min}$ are shown with dashed lines. The walls are represented by vertical colored strips.}
\label{profcase2}
\end{figure}

Figure \ref{profcase2} shows profiles corresponding to different cavity sizes, with a thin wall of size $e = 1\,{\rm cm}$ (upper panel) and a thick screened wall of size $e = 20\,{\rm cm}$ (lower panel). We find bubble profiles inside the cavity 
similar to Refs.~\cite{Brax:2013cfa, PhysRevD.87.105013}.

\begin{figure}
\includegraphics[width = 0.9\columnwidth]{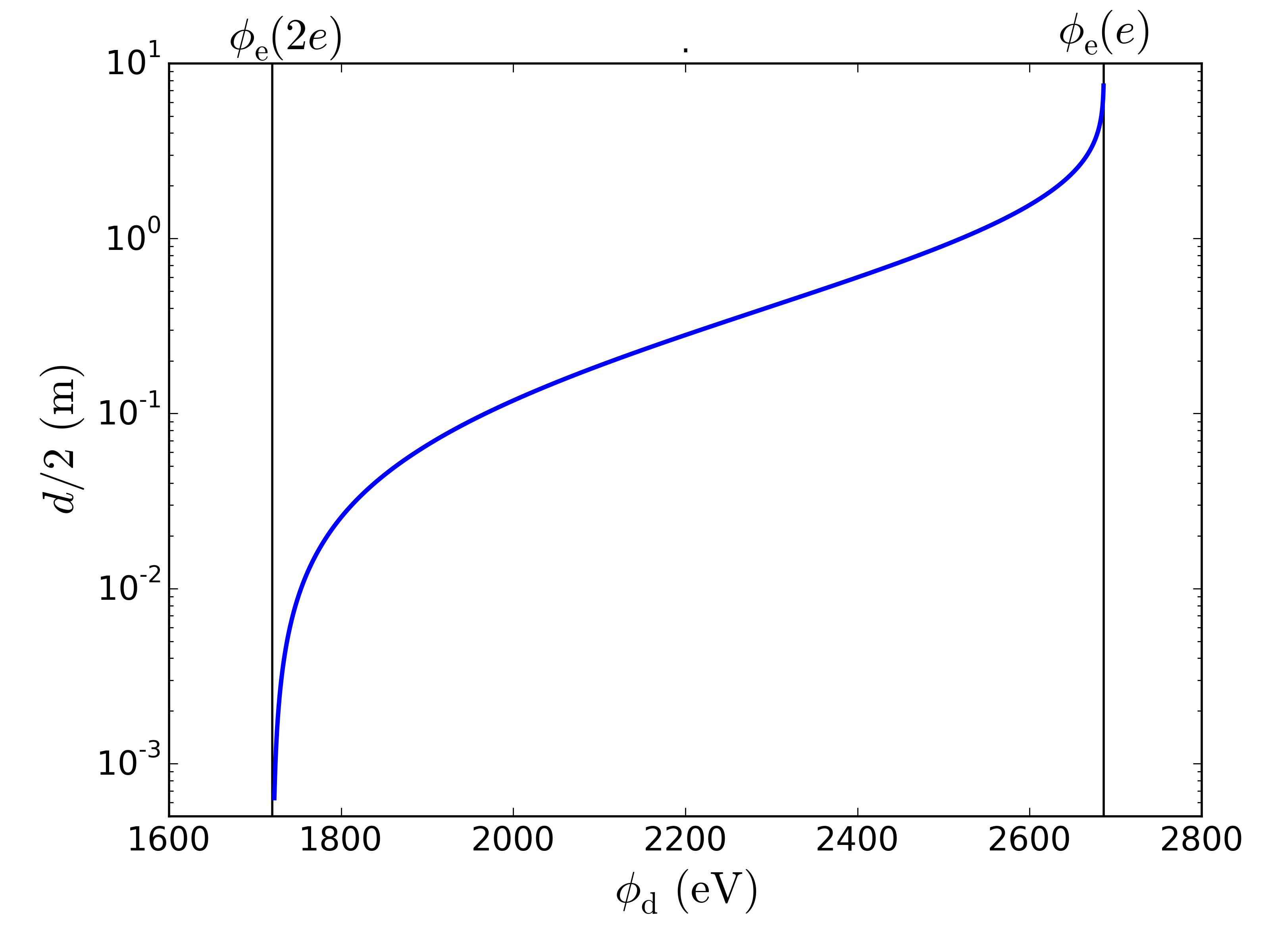}
\caption{Relation $\phi_e (d)$, for $e=1\,{\rm cm}$. $\phi_{\rm min}$ values are shown with the two black lines.}
\label{exfigreld}
\end{figure}

Figure \ref{exfigreld} shows an example of the $\phi_{\rm d}(d)$ relation with thin unscreened walls of size $e=1 \,{\rm cm}$. As for the case of a single wall, the whole interval for $d \in \mathbb{R}$ is spanned by restrained interval for $\phi_{\rm d}$, $\left[ \phi_{\rm e}(2e), \phi_{\rm e}(e) \right]$, where $\phi_{\rm e}(2e)$ corresponds to the initial condition associated to a single wall of size $2e$.

The curve for $\phi_{\rm d}(d)$  is similar to $\phi_{\rm e}(e)$ in the previous section, with two regimes. For $d \gg \lambda_{\rm c,vac}$, the field has enough space in the cavity to reach a value $\phi_0$ at its center close to the potential minimum $\phi_{\rm vac}$. In this regime and the larger the cavity is, the more we can consider the two walls as isolated, so the dynamics of the field are very similar to the one seen in Section \ref{sec:case1}. This explains why $\phi_{\rm d}$ varies very slowly with $d$, with $\phi_{\rm d} \simeq \phi_e(e)$.

On the other hand, when the cavity gets very small, the field has less and less space to evolve, such that $\phi_0$ gets smaller. In this regime, as the two walls get closer, the dynamics of the field tends to the dynamics of a single wall of thickness $2e$. This explains why small values for $d$ are obtained for $\phi_{\rm d}$ tending $\phi_e(2e)$.

\subsection{Chameleonic force in a cavity}
Having the field profiles in a cavity we can deduce the fifth force a point test mass would feel using equation \eqref{forceq}.

\begin{figure}
\includegraphics[width = 0.9\columnwidth]{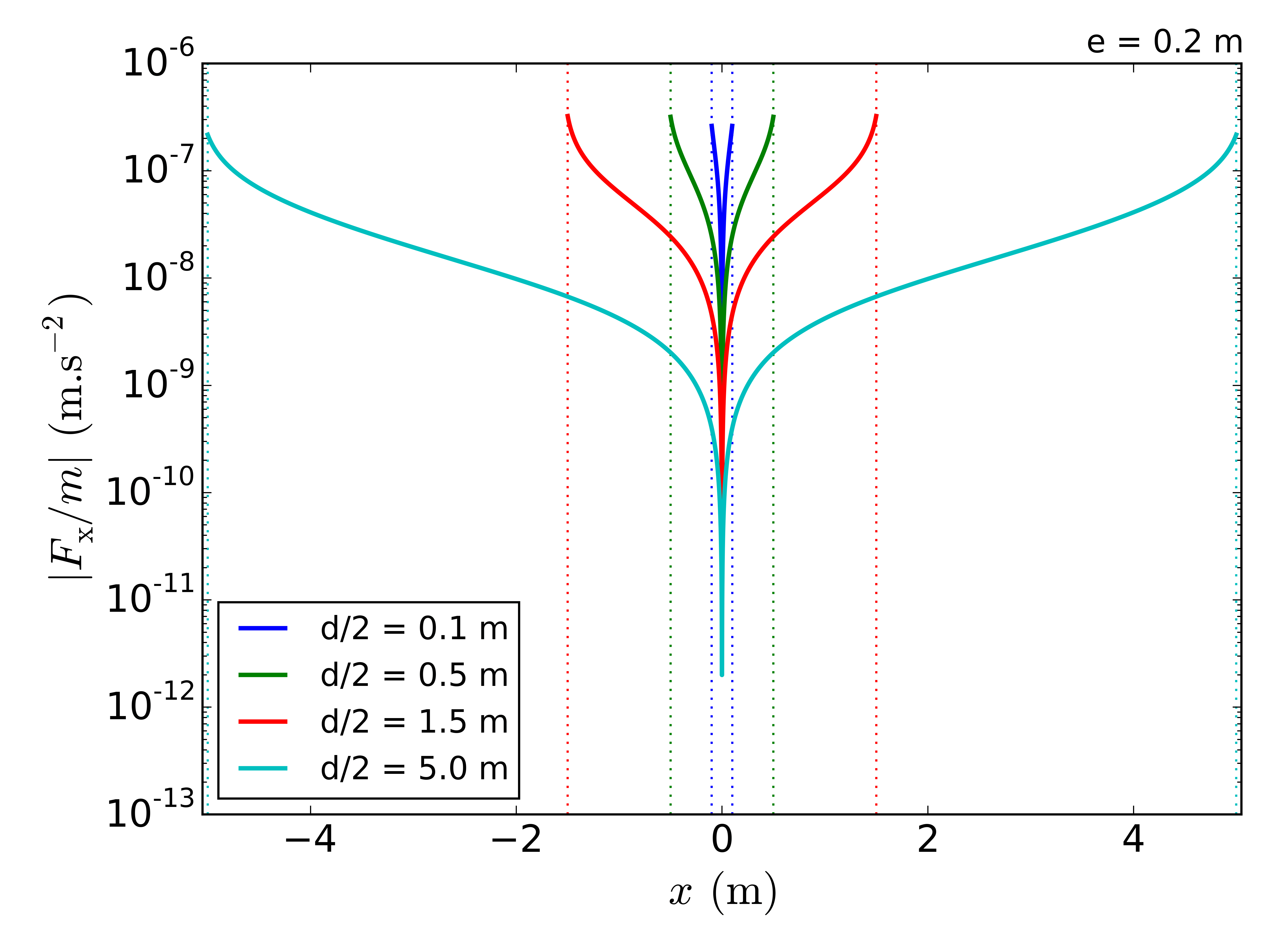}
\caption{Force experienced by a test mass for different cavity sizes, for screened walls with $e=1\, {\rm m}$.}
\label{forcevsd}
\end{figure}

\begin{figure}
\includegraphics[width = 0.9\columnwidth]{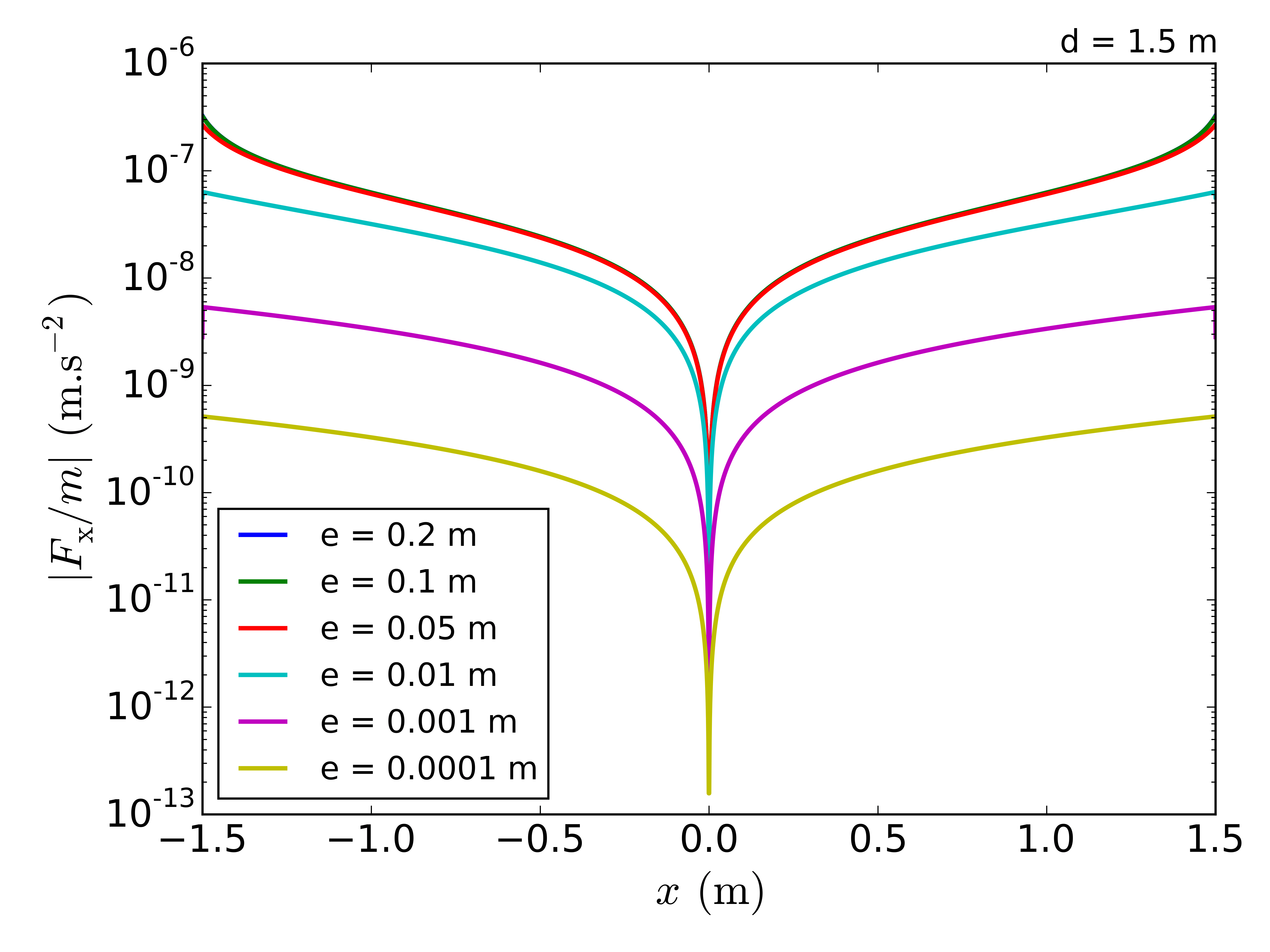}
\caption{Force experienced by a test mass for different wall sizes, for a fixed cavity size $d=1.5\, {\rm m}$.}
\label{forcevse}
\end{figure}

Figure \ref{forcevsd} shows the magnitude of the fifth force experienced by a test mass inside cavities of different sizes for a constant wall thickness as expressed by Eq.~\eqref{forceq}. This force is directed outward. The wall is chosen here to be screened with $e=0.2 \, {\rm m}$. It shows that the force profile does not vary much, but just spreads with the cavity. The maximum force value reached at the border of the cavity varies slightly.

Conversely, Figure \ref{forcevse} shows how the force profile changes as a function of the wall thickness, at constant cavity size. One sees the magnitude  of the forces  increases as walls get thicker. In agreement with previous considerations, it stops varying when the wall thickness exceeds $\lambda_{\rm c,wall} = 2.2 \,{\rm cm}$, as the screened walls isolate the inner dynamics from the outside. Thus larger  forces are expected in cavities isolated with thick walls. Nevertheless in the case of thin walls, we expect it to be overtaken by effects sourced by external objects.

\section{Cylindrical and spherical systems}
\label{sec:2D3D}
In 2D and 3D, the method previously used is no longer applicable. The chameleon's Klein-Gordon equation \eqref{ChamEq} indeed becomes in cylindrical or spherical symmetries
\begin{equation}
\frac{{\rm d}^2\phi}{{\rm d}r^2} + \frac{(D-1)}{r} \frac{{\rm d}\phi}{{\rm d}r} = n \Lambda^{n+4} \left[ \frac{1}{\phi_{\rm min}^{n+1}(\rho_{\rm mat})}-\frac{1}{\phi^{n+1}}\right],
\label{ChamEqnD}
\end{equation}
where $D$ is the dimension of the symmetry. For $D \geq 2$, the first field's derivative term prevents us from obtaining a condition on the initial field derivative by integrating once the chameleon equation. Thus we cannot follow the same scheme as previously and need to play with two initials conditions $\phi_{\rm i}$ and $\phi'_{\rm i}$.

Nevertheless, choosing now to set initial conditions at the symmetry center of the configuration is convenient, as by symmetry the derivative of the field cancels. We thus have a single parameter to determine -- the value of the field $\phi_{0}$ --, to obtain the correct profile. A dichotomy algorithm can be used to determine the correct $\phi_0$ that satisfies the correct asymptotic boundary conditions \eqref{boundarycond}. A more complex analysis of the chameleon's dynamics than the one in Section \ref{sec:chamdyn} shows that if the value of $\phi_0$ is greater (resp. weaker) than its correct value, the field will systematically asymptotically diverge positively (resp. negatively). Then solving the field for some $\phi_0$, we can evaluate if at some large distance far greater than  $\lambda_{\rm c,vac}$ from the considered system the field is greater or lesser than $\phi_{\rm vac}$, and then adjust $\phi_0$ as a dichotomy and reproduce the same procedure.

\begin{figure}
\includegraphics[width = 1\columnwidth]{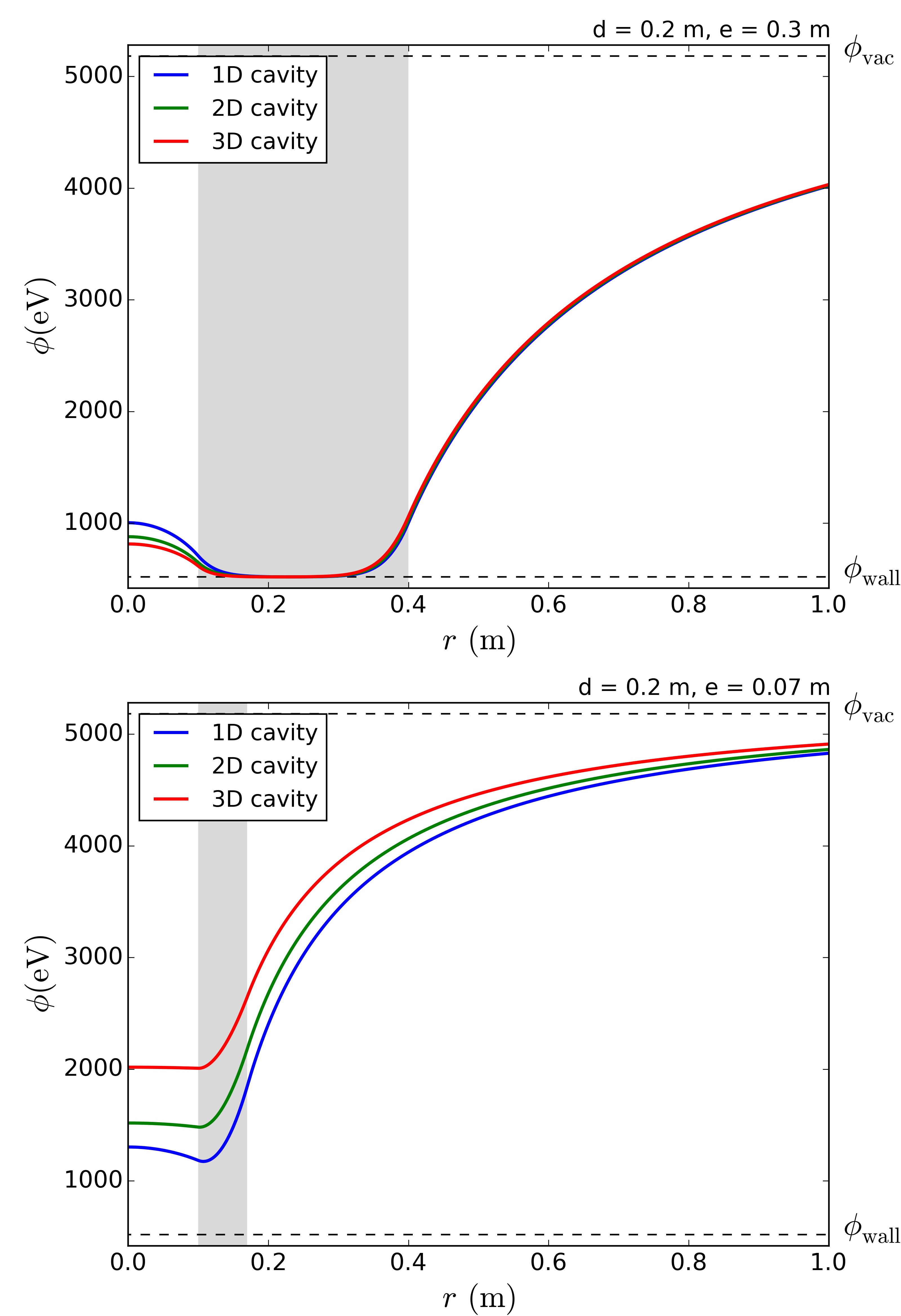}
\caption{Field profiles for 1D, 2D and 3D cavity in the case of screened walls of thickness $e = 30\,{\rm cm}$ and of un-screened walls of thickness $e = 7\,{\rm cm}$. $\phi_{\rm min}$ are shown with dashed lines. The wall is shown with the grey region.}
\label{profcavnD}
\end{figure}

This  converges rapidly toward the correct profile. It is important to note that the symmetry center being the origin of the coordinate system $r = 0$, we cannot impose initial conditions at this point as the second term in Eq.~\eqref{ChamEqnD} diverges numerically due to its dependence in $r$. We instead impose them very close to $r=0$ with $\phi_{\rm i} = \phi_0$ and $\phi'_{\rm i} = 0$. This should lead to an error on the obtained field.  The fields obtained in a 1D cavity with this method agree with the fields obtained with the previous method to less than 0.1\%.

\subsection{Cylindrical and spherical cavity}
\label{sec:cyl}

Analogously to Section \ref{sec:case2}, the case of a cavity in 2D and 3D are respectively the infinitely extended cylinder and the empty sphere. We still denote here by $d$ the diameter of the cavity and $e$ the wall thickness.

The radial profiles in such cases are very similar to the 1D case. For equal cavity size, the effect of cylindrical and spherical symmetry decreases the values reached in the cavity. Figure \ref{profcavnD} shows examples of radial profiles for both 1D, 2D and 3D cavities in the case of screened walls and non-screened walls.

When the wall is screened the nature of the cavity does not affect the field outside. The field tends to reach lower values in the cavity for larger  cavity sizes, leading to weaker force. When the wall is not screened the behavior gets inverted and the size of the cavity has an impact on the exterior field.

\subsection{$\bm{\phi_0}$ variation}
\label{sec:phi0var}
\begin{figure}
\includegraphics[width = 0.9\columnwidth]{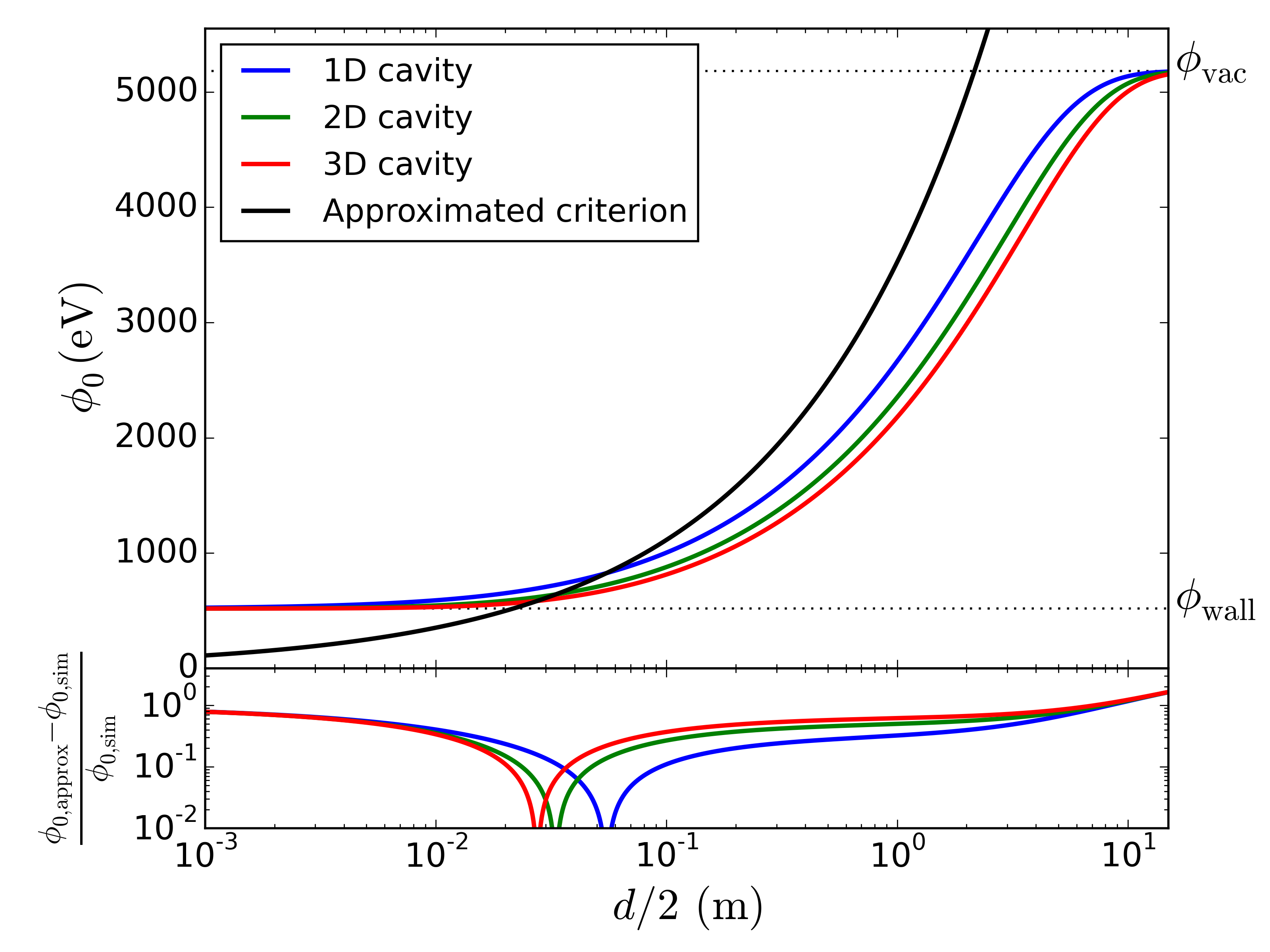}
\caption{Central value of the field in the cavity as a function of the cavity size $d$ for screened walls of thickness $e=0.1\,{\rm m}$. The colored lines correspond to 1D, 2D and 3D cavities. The black line is an approximated estimation from \cite{khoury_chameleon_2004}. $\phi_{\rm min}$ values are shown with the two dotted lines. The lower panel shows the relative difference between the two curves.}
\label{phimidcav}
\end{figure}

As for the 1D cavity, the larger the cavity the larger the value $\phi_0$ reached by the field at the center of the cavity. In the literature (e.g. Ref.~\cite{khoury_chameleon_2004} for a sphere or Refs.~\cite{brax_testing_2007, BraxReview} for a cylinder), this value was expected to be that of the field whose mass matches the radius of the cavity, i.e. solving
\begin{equation}
\frac{d}{2} = m^{-1}(\phi_0) = \frac{1}{\sqrt{V''(\phi_0)}}.
\end{equation}

In Figure \ref{phimidcav} the value of $\phi_0$ obtained with this approximate criterion is compared to the actual value given by these simulations for a 1D, 2D and 3D cavity. All curves have the same global monotony. Nevertheless whereas simulations shows that $\phi_0 \in [\phi_{\rm wall}, \phi_{\rm vac}]$, the approximated criterion does not give a bounded range for $\phi_0$. The comparison of the curves shows that they mainly diverge by 100\%, such that the approximated criterion turns out to be very weak.

\section{Applications and discussion}
\label{sec:appli}
\subsection{Chameleonic Casimir-like force}

The one-dimensional configuration in Section \ref{sec:case2} is similar to the typical experimental set-up in Casimir effect measurements in which two nearby plates would experience a force of quantum origin \cite{sushkov_observation_2011, lambrecht_casimir_2012}. In the case of the chameleon field, one expects an extra effect that would add-up to the conventional Casimir force. In both cases, the force between the walls is attractive. The walls play the role of the plates, and the effect originates from the fact that the field in the walls is not symmetrical so that its gradient does not cancel. The global behavior of the force as a function of the distance between the walls was computed with an approximate analytic model in Ref.~\cite{brax_detecting_2007}.

The force a wall feels can be computed by integrating the gradient of the field over the whole wall. Knowing the profile associated to a two-wall configuration, a 1D integration gives a pressure
\begin{equation}
\begin{split}
F_{s}&= - {\rm c}^2 \frac{\beta}{M_{\rm Pl}} \int_{\frac{d}{2}}^{\frac{d}{2} + e}\ \nabla_x\phi \, \rho_{\rm wall} \, {\rm d}x \\
&= - {\rm c}^2 \frac{\beta}{M_{\rm Pl}} \rho_{\rm wall} \, [\phi_{\rm d} - \phi(d/2)].
\end{split}
\end{equation}

\begin{figure}
\includegraphics[width = 0.9\columnwidth]{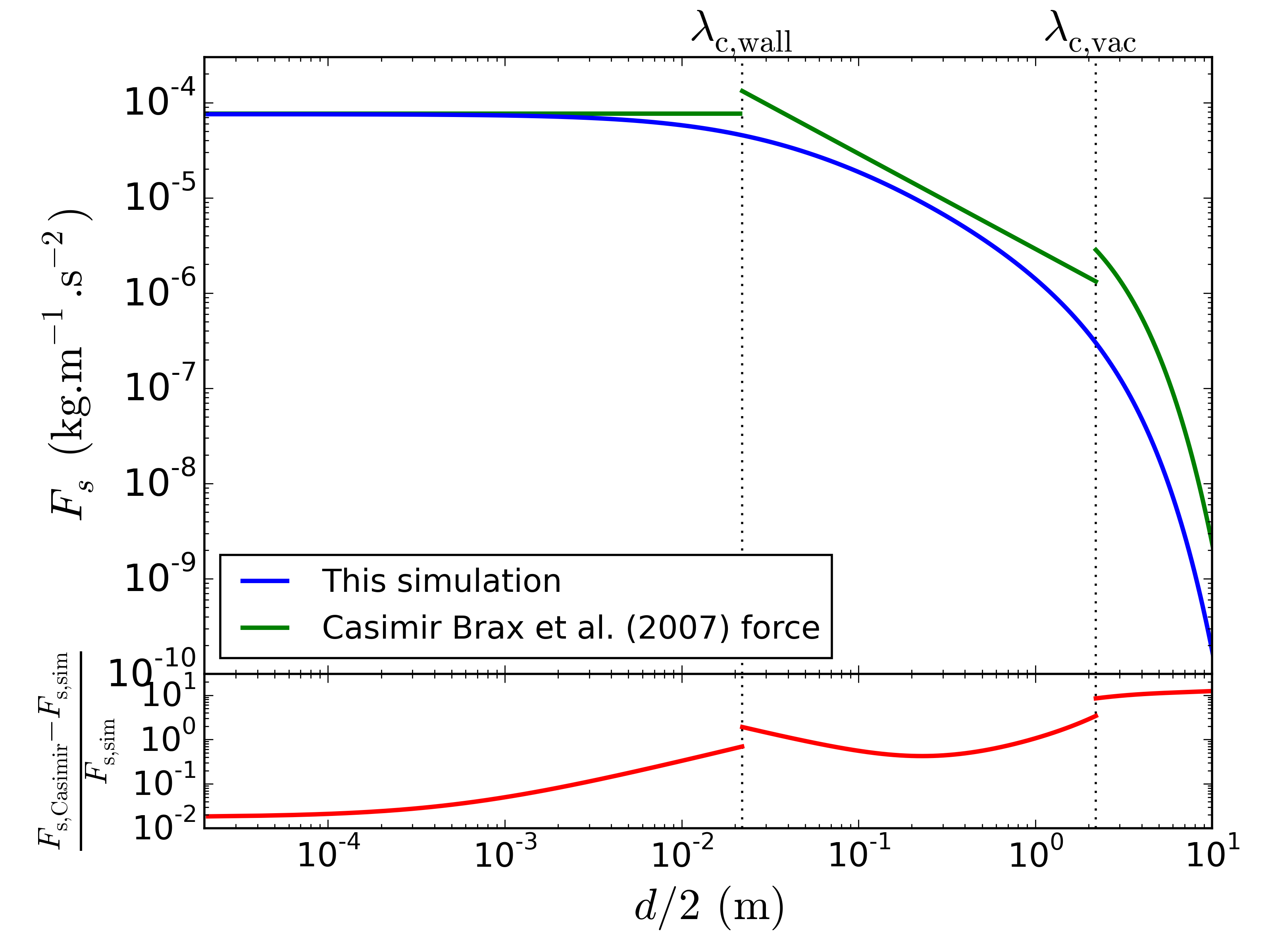}
\caption{Upper panel: chameleon-originated Casimir force as a function of the separation $d$ of the walls for screened walls with $e=0.2 \,{\rm m}$. The blue curve is the result of this simulation. The green lines come from an analytical model from Ref.~\cite{brax_detecting_2007}. Lower panel: relative difference. Dotted lines show $\lambda_{\rm c,vac}$ and $\lambda_{\rm c,wall}$. }
\label{casimirforce}
\end{figure}

Figure \ref{casimirforce} shows the evolution of this pressure in the case of our simulations and in the case of Ref.~\cite{brax_detecting_2007}, as a function of the separation of  the walls. Both curves have the same global behavior with a plateau for very close walls, and with an exponential suppression for separation greater than $\lambda_{\rm c,vac}$. This latter behavior is consistent with Section \ref{sec:phi0var}, as we saw that for large separation the walls can be considered as isolated, so the field tends to the symmetrical case of Section \ref{sec:case1}.

Despite their similar behavior, the two curves do not match perfectly. For small separation they agree within a few percent. In the intermediate regime $\lambda_{\rm c,wall} < \frac{d}{2} < \lambda_{\rm c,vac}$, they diverge by a few tenths of percent, and for the larger separation they diverge more severely. The force we find being weaker, this might slightly relax current Casimir measurement constraint on the chameleon \cite{brax_detecting_2007}.

\begin{figure*}
\includegraphics[width = 0.7\textwidth]{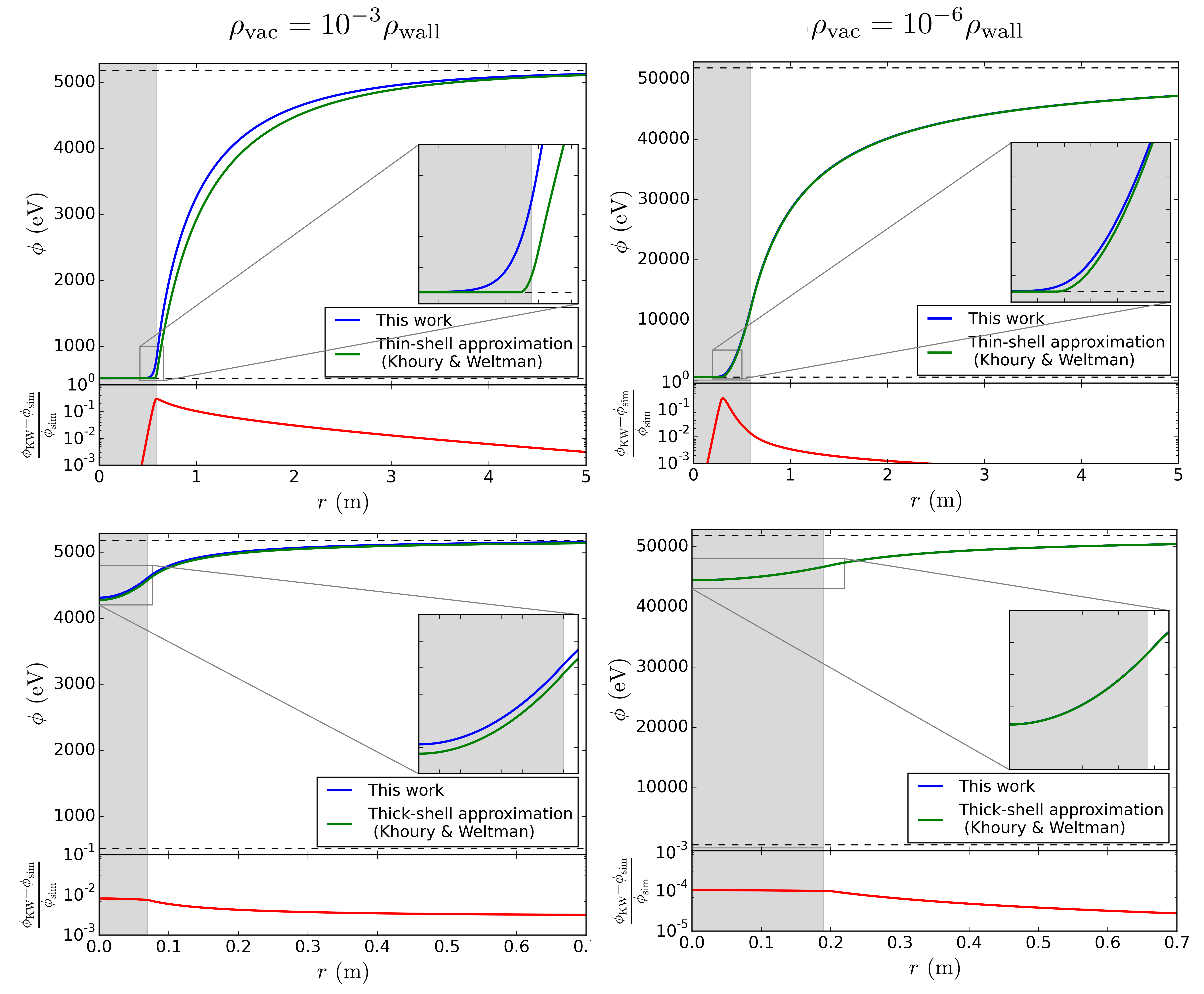}
\caption{Radial field profiles of a ball embedded in vacuum. The shaded zone corresponds to the inside of the ball. These simulations are compared to models of Ref.~\citep{khoury_chameleon_2004} in the thin-shell regime (upper panels) and thick-shell regime (lower panels). The left and right panels correspond to different matter contrast between the ball and the vacuum. $\phi_{\rm min}$ values are shown with dashed lines.}
\label{phiballKW}
\end{figure*}

\subsection{Thin and thick-shell approximations of a ball}

Another important case is the spherical uniform ball. In the chameleon's original article \citep{khoury_chameleon_2004}, the profile around a ball was approximated in two extremal regimes: the thick-shell regime in which the ball is too small for the field to reach the minimum of the potential in the ball; and the thin shell regime in which the ball is large enough for the field to remain mainly at the minimum of the potential in most of the ball. Our simulations can provide the field around a ball in any regime.

Figure \ref{phiballKW} compares our simulations with the thin-shell and thick-shell approximations, with different contrasts between the vacuum and the ball density, then different $\phi_{\rm vac}$. In the thick-shell regime, our simulation and the thick-shell model are in very good agreement to less than a percent when the density contrast is low. When the density contrast is larger, the agreement gets even better to less than 0.01 \%. In the thin-shell regime, the two profiles agrees to a few percent except inside a zone around the interface between the ball and the vacuum, where they only agree to a few tenths of percent. This comes mainly from a difference in the skin depth of the wall on which the field varies. For higher density contrast, the agreement is globally better, with still a slight difference where the field starts to vary inside the ball.

We can therefore assume the models to be globally accurate except inside the ball in the thin-shell regime.  In between the two regimes, when the ball has an intermediate radius, these two models are less accurate, particularly for low density contrasts.

\vspace{0.5cm}
\subsection{Radial drift of test masses in a cylinder}
\begin{figure}
\includegraphics[width = 0.9\columnwidth]{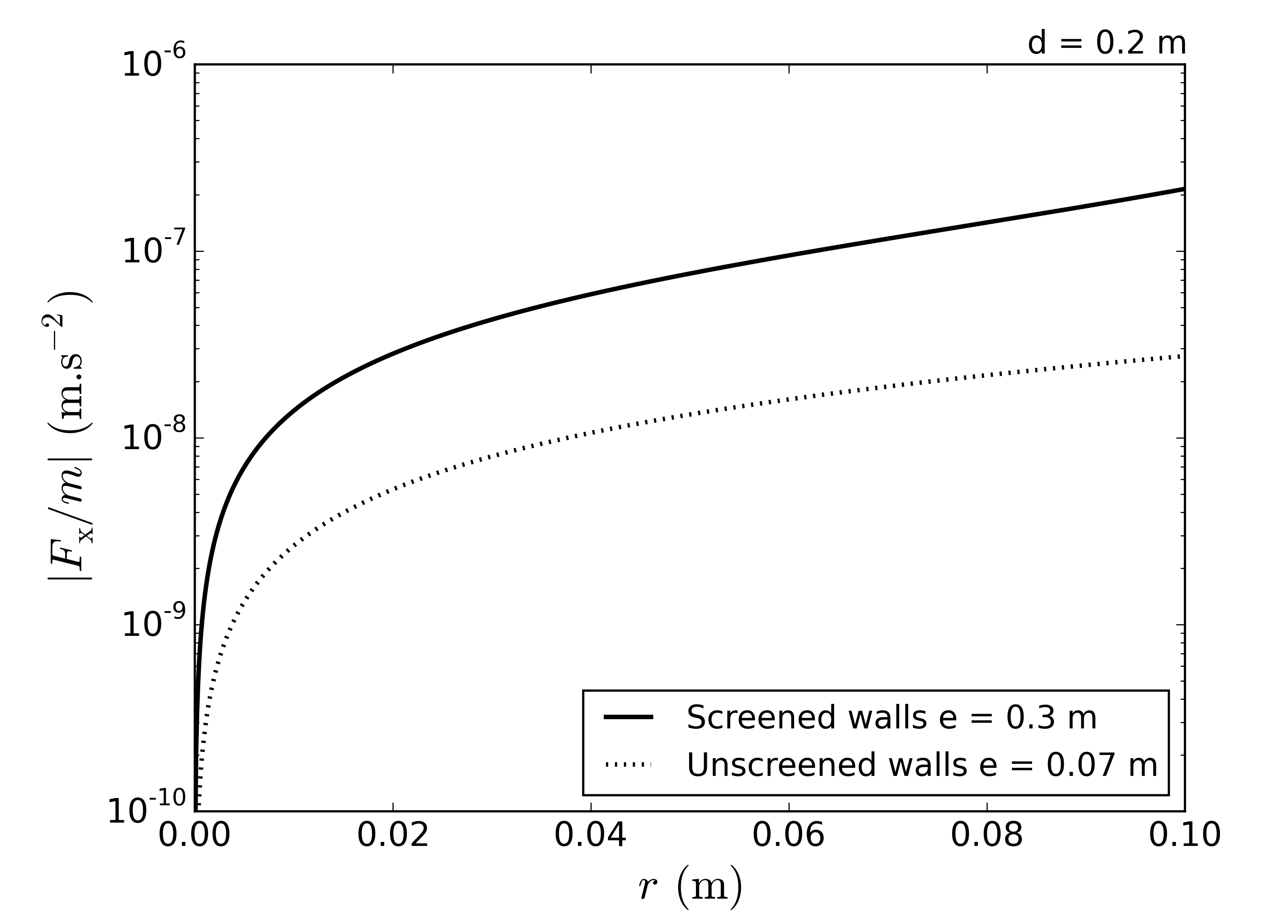}
\includegraphics[width = 0.9\columnwidth]{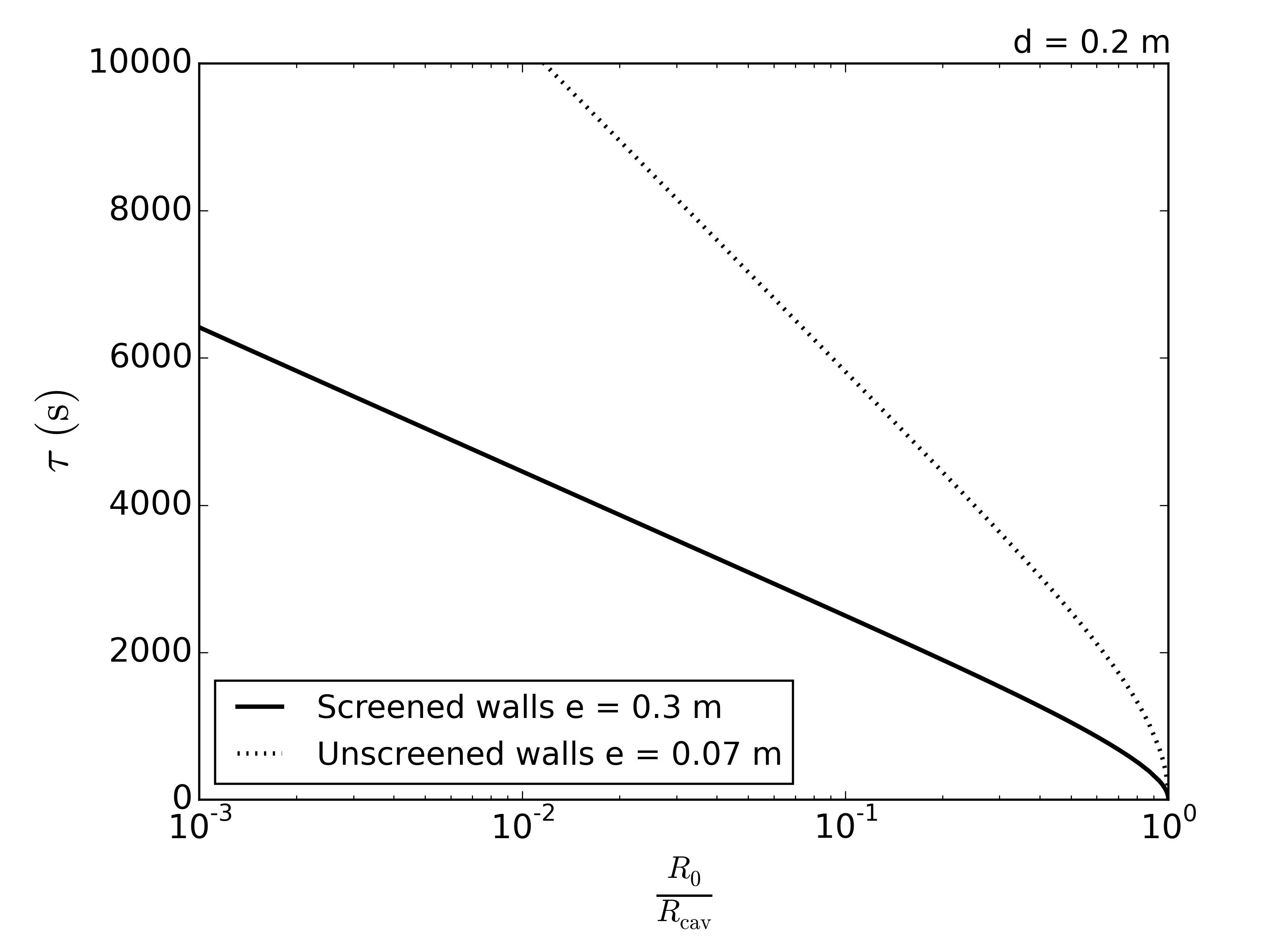}
\includegraphics[width = 0.9\columnwidth]{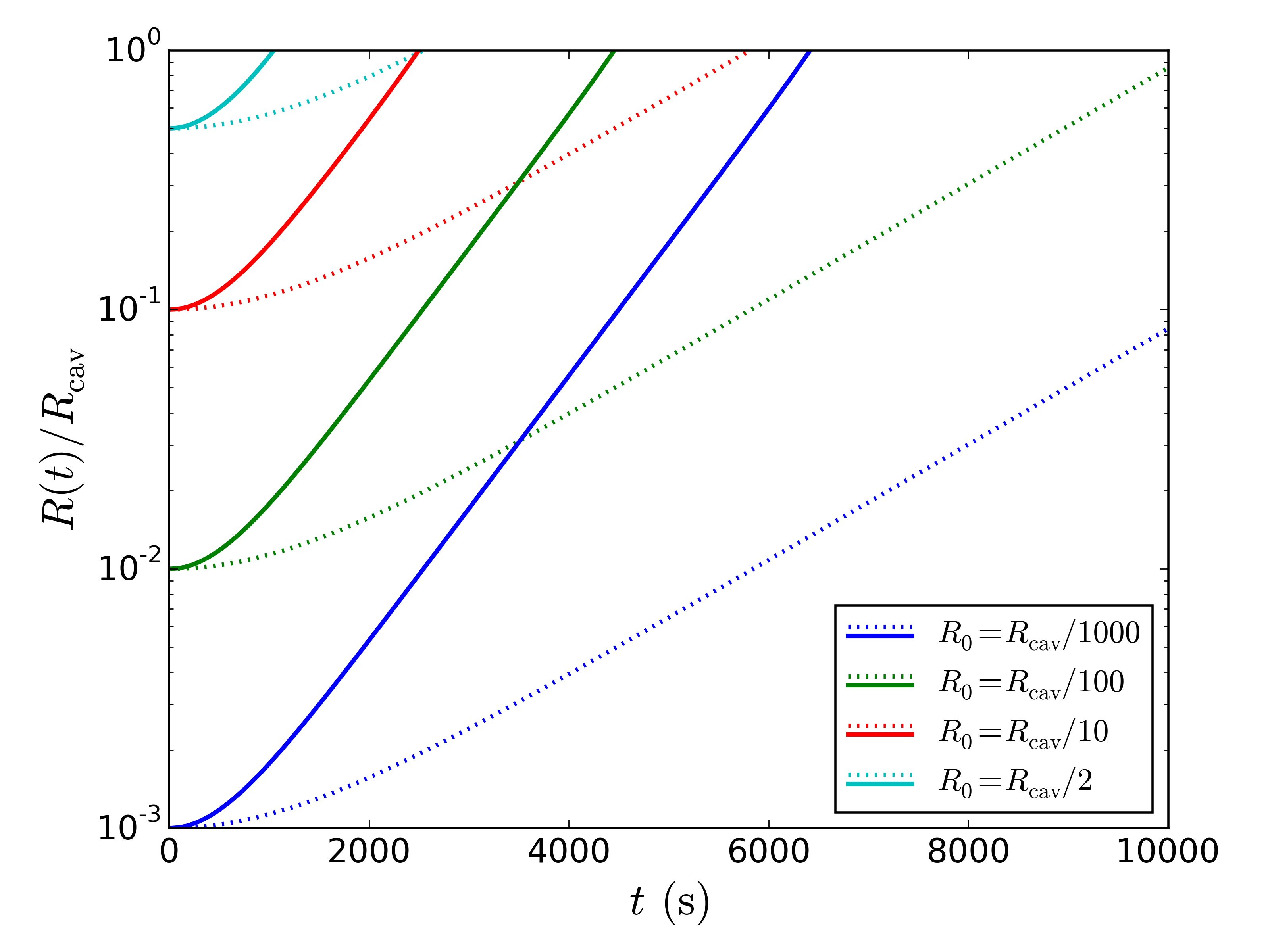}
\caption{Upper panel: fifth force associated to the cylindrical cavity in Fig.~\ref{profcavnD}. Middle panel: total drift time from some initial position $R_0$ to the border of the cavity $R_{\rm cav}$. Lower panel: radial position of atoms $R$ as they are drifting with time $t$ for different initial positions. Solid lines correspond to screened walls ($e = 0.3\,{\rm m}$) and dotted lines to unscreened walls ($e = 0.07\,{\rm m}$).}
\label{driftfigs}
\end{figure}

As shown above, the chameleon inside a cavity creates a radial outward force that affects test masses (like atoms). For instance, the profile of the force created in the cylindrical case of Fig.~\ref{profcavnD} is shown in the upper panel of Fig.~\ref{driftfigs}. This force may affect any experiment based on monitoring the trajectory of atoms inside a cylindrical cavity \cite{Llinares:2018mzl}, even if measuring it is not the primary objective of the experiment (in which case it should be considered as a source of systematic uncertainty).

Let us consider an experiment where atoms (test masses) are dropped at a distance $R_0$ from the main symmetry axis of the cylinder (either alongside the axis, or radially): the atoms will experience a outward radial drift, with a drift rate depending on the parameters ($\beta,\Lambda, n$) of the model. For instance, in the screened cylindrical case of Fig.~\ref{profcavnD}, if they are dropped with a null velocity at $R_{\rm vac}/10$, the atoms will reach the border of the cavity in $2498~{\rm s}$.
The middle panel of Fig.~\ref{driftfigs} shows the total drift time for the atoms to reach $R_{\rm cav}$ as a function of $R_0$. In the unscreened case, as the profile is flatter, the force is weaker than in the screened case, so that the drift time is typically longer. Trivially, the smaller $R_0$, the longer the drift time. The lower panel shows the evolution with time $t$ of the radial position $R$ of atoms in the cavity for different initial positions $R_0$.

In more realistic setups, the drift should be estimated in view of typical integration times, as it may become non-negligible, even in experiments not specifically looking for a chameleon inside the cavity. For instance, we could conceive of an experiment where the motion of atoms under the influence of the Earth gravity field is measured. If the chameleon force inside the cavity is strong enough to impart a detectable drift on the atoms, it should be considered as a systematics (though its detection would be a significant breakthrough). Another typical case is where we drop two types of atoms, e.g. to test the equivalence principle in the Earth gravity field; if the chameleon coupling $\beta$ is not universal, then the chameleon inside the cavity will make the atoms drift differentially, thereby mimicking a violation of the equivalence principle, though it would be considered as a systematic uncertainty on the main measurement.

\subsection{Nested cylinders: toward the MICROSCOPE configuration}
\begin{figure}
\includegraphics[width = 0.9\columnwidth]{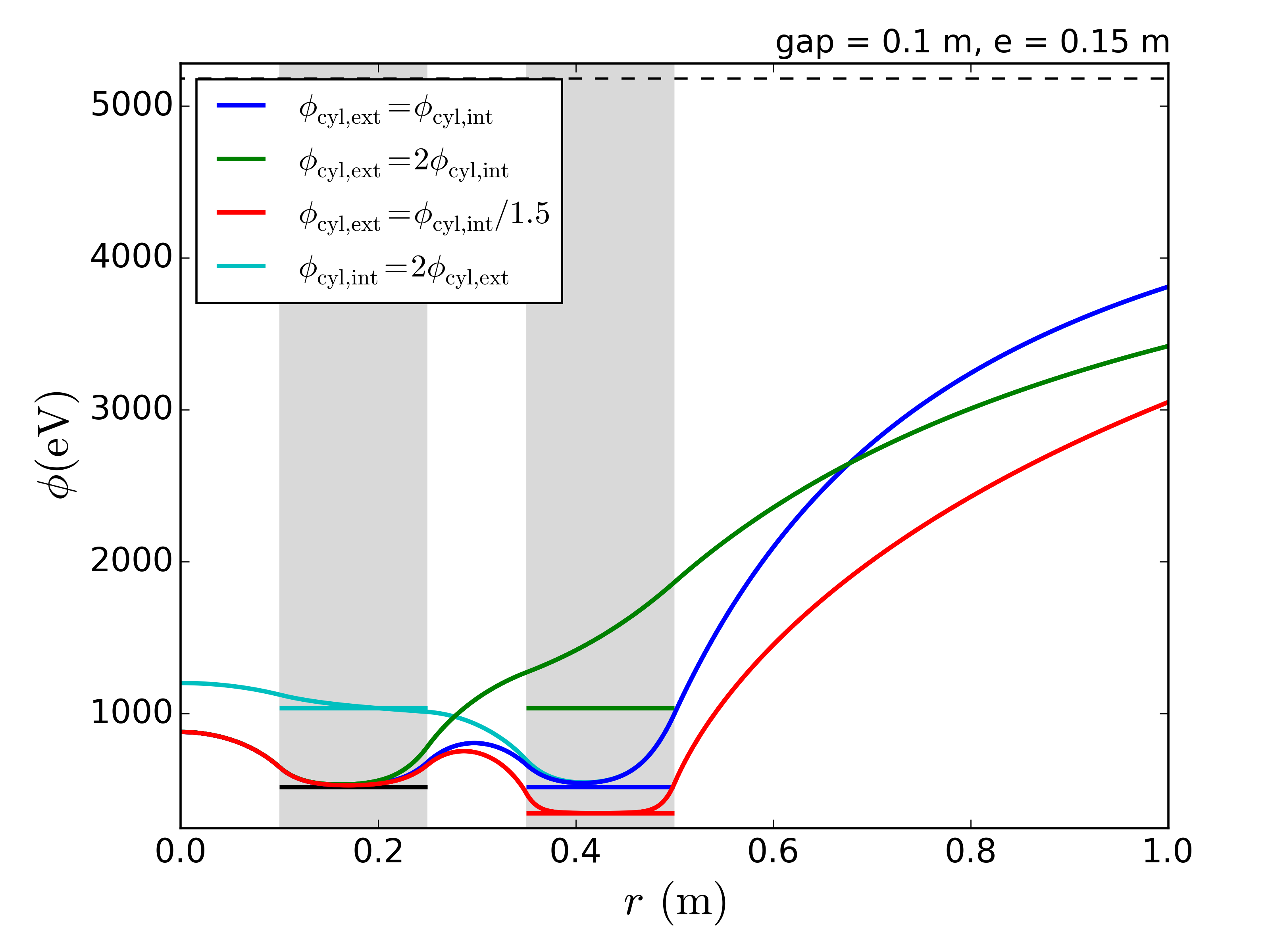}
\caption{Radial profiles for 2 nested cylinders of thickness $e$ and different matter densities to which corresponds different values of $\phi_{\rm min}$. These $\phi_{\rm min}$ values are represented with the horizontal segments. Cylinders are delimited by the shaded regions and separated by distance $gap$.}	
\label{phinested}
\end{figure}

Our computation generalizes to more complex configurations, such as the case of nested infinite cylinders. Figure \ref{phinested} compares different profiles for two nested cylinders of same density or different matter densities. Consistently with our 1D study, the nature of the outer cylinder has no influence on the profile inside the inner cylinder  when the cylinders are screened. Besides, in the inter-cylinder empty space, atoms experience a drift similar to the one discussed previously. But whereas in the cylindrical cavity, a change of direction  in the force occurs at the center of the cavity, here it no longer occurs at the middle of the empty space but at some other location -- see the maxima of the field -- that depends on the densities of the cylinders and on the parameters of the model. This change of direction can even disappear, as in the green line. Then different signatures are expected for different cylinders' features and chameleon parameters.

\begin{figure}
\includegraphics[width = 0.9\columnwidth]{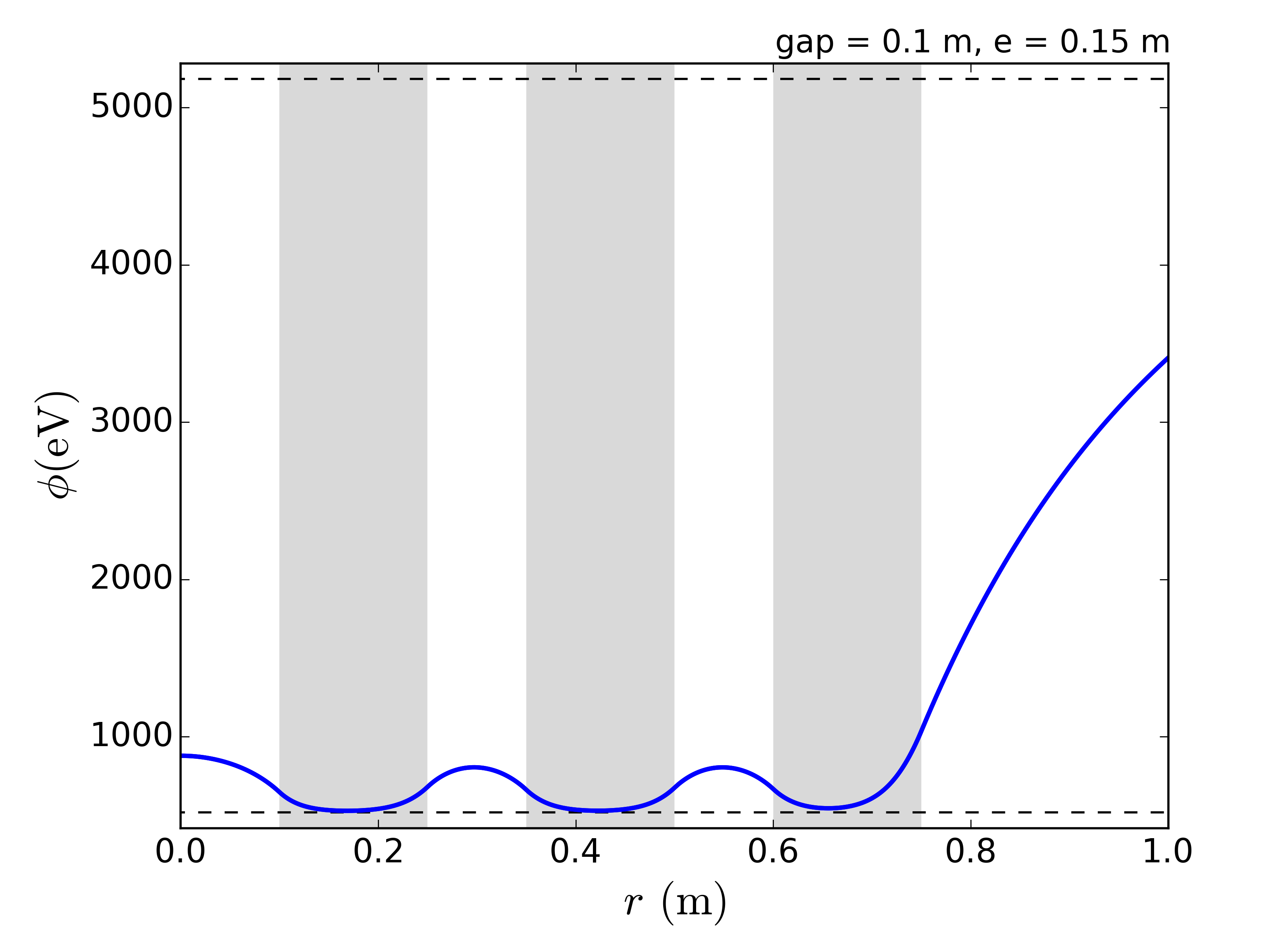}
\caption{Radial profile for 3 nested cylinders of thickness $e$ same matter density. Cylinders are delimited by the shaded regions and separated by a distance $gap$. The $\phi_{\rm min}$ values are represented with the horizontal segments.}	
\label{phi3nested}
\end{figure}

Figure \ref{phi3nested} shows the radial profile obtained for 3 nested cylinders. This configuration is similar to the MICROSCOPE experiment design in which cylindrical test masses are nested in cylindrical sensors. The middle cylinder experience a chameleonic fifth force from the cylinders, nevertheless when integrated over the whole cylinder it vanishes due to the cylindrical symmetry. We expect a force to appear when the symmetry is broken, e.g. when one of the cylinders is not perfectly centered. While this would require more intricate computations, described in a follow-up article, we can estimate the magnitude of such a force. To that purpose, we consider the force exerted on a cylindrical element (of opening angle ${\rm d} \theta$ and height ${\rm d} l$) of a cylinder. Here, in the case of Fig.~\ref{phi3nested}, $\frac{{\rm d} F}{{\rm d} \theta ~{\rm d} l} = 1.1 \times 10^{-5} \,{\rm N.m^{-1}.rad^{-1}}$ and the force is directed toward the center. We expect the total force in a decentered configuration to be of the same order of magnitude up to geometry factor.

\begin{figure}
\includegraphics[width = 0.9\columnwidth]{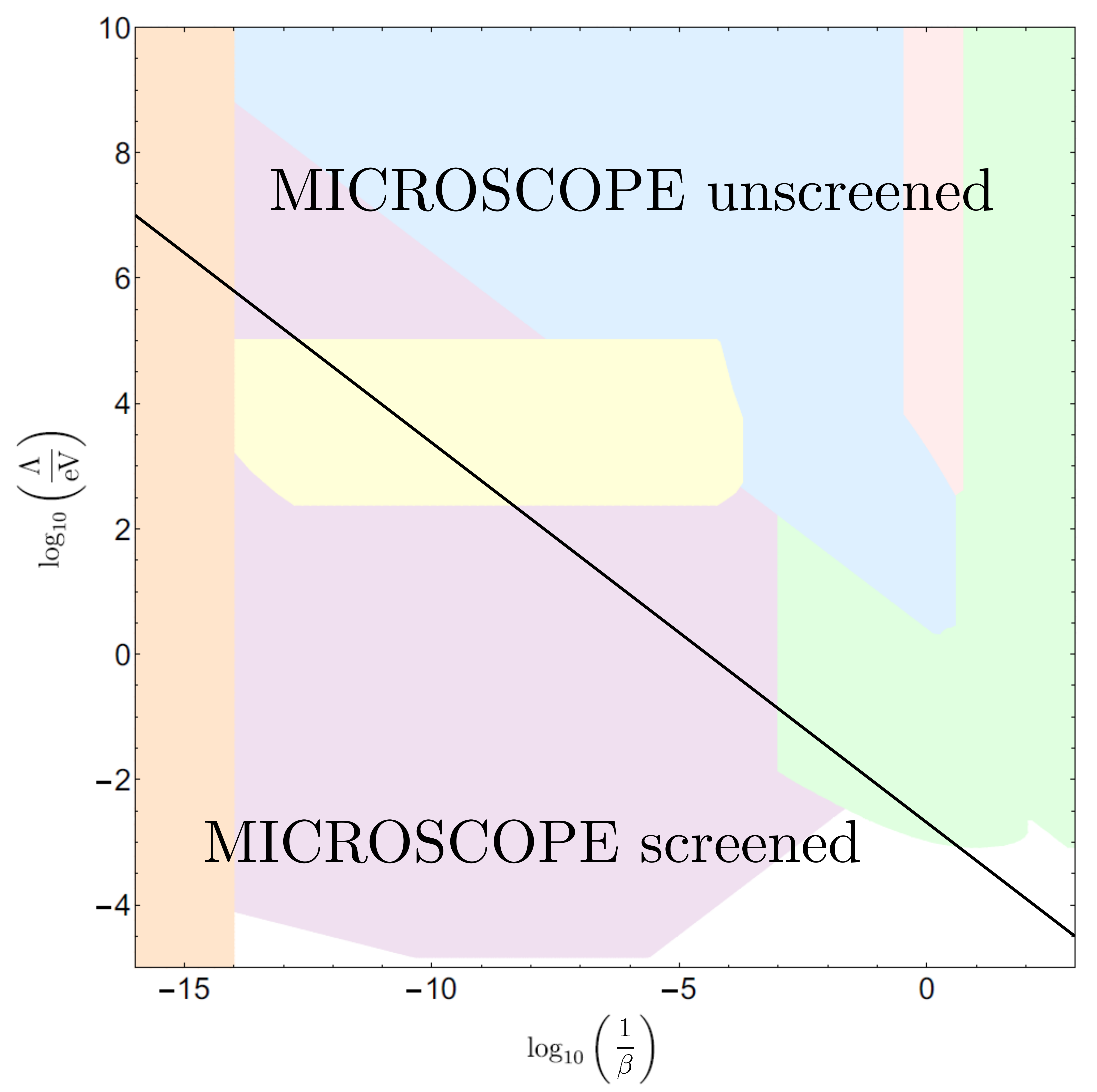}
\caption{Chameleon's parameter space adapted from Refs.~\cite{burrage_tests_2018,BraxReview}. The black line corresponds to parameters for which $100 \,\lambda_{\rm c, shield} = e_{\rm shield} / 2$ and delimits two regimes whether the MICROSCOPE set-up is screened or not. Colored region corresponds to current constraints from other experiments as atomic interferometry (purple, \cite{jaffe_testing_2017}), E\"ot-Wash (green, \citep{upadhye_dark_2012}), Casimir effect measurements (yellow, \cite{brax_detecting_2007}), astrophysics tests (blue, \cite{Jain_2013}) and lensing (pink, \cite{burrage_tests_2018}), microspheres (blue line, \cite{PhysRevLett.117.101101}) or precision atomic tests (orange, \cite{PhysRevD.83.035020}).}	
\label{MICparam}
\end{figure}

In Ref.~\cite{khoury_chameleon_2004} , it was claimed that MICROSCOPE could detect a clear violation of the weak equivalence principle from the chameleon field sourced by the Earth. However, the screening due to the experimental set-up itself was neglected. The MICROSCOPE set-up is actually enclosed in a shield of thickness $e_{\rm shield} \simeq 1\, {\rm cm}$. Using the screening criterion of Section \ref{sec:screening}, we show in Figure \ref{MICparam} that the chameleon parameter space (for $n=1$) is divided in two regions: above the black line (that shows where $100 \,\lambda_{\rm c, shield} = e_{\rm shield} / 2$,  where $\lambda_{\rm c, shield}$ stands for the Compton wavelength associated to the shield's density), MICROSCOPE is not screened, though it is screened below the line. Thus, no violation of the WEP can be expected below the line, while some could still be expected above it. The colored regions in Fig.~\ref{MICparam} correspond to regions already experimentally excluded \cite{burrage_tests_2018}. It is then clear that the constraining potential of MICROSCOPE is much less that anticipated. Only in a small region could it improve our current knowledge on the chameleon. This will be the subject of future work where the effect of Earth on the chameleon profile should be included.
   
\section{Conclusion}
\label{sec:concl}
In this article, we treated the chameleon problem of solving for the scalar field's profile by paying special attention to the boundary conditions. We found that it is possible to deal numerically with this problem without any approximation. Our approach is to consider a matter system embedded in a background environment. We first considered 1D symmetrical systems. We treated the case of a single wall and of a cavity modeled as  two separated walls. We determined a refined criterion guaranteeing that screening occurs within a cavity. For instance we checked that we can safely consider that the field reaches its minimum inside a matter wall, as long as the wall thickness exceeds one hundred times the Compton wavelength associated to the wall matter. In this case we can consider that such a wall would screen the field. In the case of a cavity, we computed the profiles of forces that test masses would experience inside a cavity. We also computed the Casimir-like force and found discrepancies  with analytic approximations in the literature.  We then explored 2D and 3D symmetrical geometries. The case of a ball has been compared to the thin shell and thick shell models from Ref.~\citep{khoury_chameleon_2004}. We found it to be in very good agreement except in the region close to the ball boundary. In a cylindrical cavity, we  studied how point masses like atoms could experience a drift between the cylinders which may either lead to an experimental method of detecting chameleons or create a new source of systematic uncertainty in future experiments.

Finally, we treated the case of nested cylinders of different matter densities suited to the set-up of the MICROSCOPE mission. Despite the symmetry considered here, leading to a null force experienced by the cylinders, we provided an estimate of the magnitude of the force when the symmetry is broken. This effect will be explored by simulating disymmetrical configurations in a follow-up article. Moreover, our analysis challenges the previous claim on the ability of space experiment to detect chameleon-originated violations of the weak equivalence principle sourced by the Earth \cite{khoury_chameleon_2004a,khoury_chameleon_2004}. Using the refined screening criterion for cavities, we deduced that for a large region of the parameter space, such effect would be screened by the experimental set-up.  For the remaining region, the Earth should be included in these simulations. This will be the subject of a forthcoming article.

\section*{Acknowledgment}
We thank Manuel Rodrigues, Gilles M\'etris and Pierre Touboul for useful discussions and technical information about the MICROSCOPE instrument. We thank the members of the MICROSCOPE Science Working Group for allowing us to start this project and encouraging us to pursue it. We acknowledge the financial support of CNES through the APR program (``GMscope+'' project). MPB is supported by a CNES/ONERA PhD grant. This work uses technical details of the T-SAGE instrument, installed on the CNES-ESA-ONERA-CNRS-OCA-DLR-ZARM MICROSCOPE mission. This work is supported in part by the EU Horizon 2020 research and innovation programme under the Marie-Sklodowska grant No. 690575. This article is based upon work related to the COST Action CA15117 (CANTATA) supported by COST (European Cooperation in Science and Technology).
\bibliographystyle{ieeetr}
\bibliography{Bib}

\begin{thebibliography}{10}

\bibitem{will_confrontation_2014}
C.~M. Will, ``The {Confrontation} between {General} {Relativity} and
  {Experiment},'' {\em Living Rev. Relativ.}, vol.~17, p.~4, Dec. 2014.

\bibitem{Ishak2018}
M.~Ishak, ``Testing general relativity in cosmology,'' {\em Living Reviews in
  Relativity}, vol.~22, p.~1, Dec 2018.

\bibitem{PRLgw2016}
{LIGO Scientific Collaboration and Virgo Collaboration}, ``Observation of
  gravitational waves from a binary black hole merger,'' {\em Phys. Rev.
  Lett.}, vol.~116, p.~061102, Feb 2016.

\bibitem{abbott_multi-messenger_2017}
{LIGO Scientific Collaboration and Virgo Collaboration}, ``Multi-messenger
  {Observations} of a {Binary} {Neutron} {Star} {Merger},'' {\em ApJL},
  vol.~848, no.~2, p.~L12, 2017.

\bibitem{Planck2018}
{Planck Collaboration}, Y.~{Akrami}, F.~{Arroja}, M.~{Ashdown}, J.~{Aumont},
  C.~{Baccigalupi}, M.~{Ballardini}, A.~J. {Banday}, R.~B. {Barreiro}, and
  N.~{Bartolo}, ``{Planck 2018 results. I. Overview and the cosmological legacy
  of Planck},'' {\em arXiv e-prints}, p.~arXiv:1807.06205, Jul 2018.

\bibitem{PrimCosmo}
P.~Peter and J.-P. Uzan, {\em {Primordial cosmology}}.
\newblock Oxford Graduate Texts, Oxford: Oxford Univ. Press, 2009.

\bibitem{2013arXiv1309.5389J}
B.~{Jain}, A.~{Joyce}, R.~{Thompson}, A.~{Upadhye}, J.~{Battat}, P.~{Brax},
  A.-C. {Davis}, C.~{de Rham}, S.~{Dodelson}, A.~{Erickcek}, G.~{Gabadadze},
  W.~{Hu}, L.~{Hui}, D.~{Huterer}, M.~{Kamionkowski}, J.~{Khoury}, K.~{Koyama},
  B.~{Li}, E.~{Linder}, F.~{Schmidt}, R.~{Scoccimarro}, G.~{Starkman},
  C.~{Stubbs}, M.~{Takada}, A.~{Tolley}, M.~{Trodden}, J.-P. {Uzan},
  V.~{Vikram}, A.~{Weltman}, M.~{Wyman}, D.~{Zaritsky}, and G.~{Zhao}, ``{Novel
  Probes of Gravity and Dark Energy},'' {\em arXiv e-prints},
  p.~arXiv:1309.5389, Sep 2013.

\bibitem{test_STtheory_gilles}
P.~C.~C. Freire, N.~Wex, G.~Esposito-Farèse, J.~P.~W. Verbiest, M.~Bailes,
  B.~A. Jacoby, M.~Kramer, I.~H. Stairs, J.~Antoniadis, and G.~H. Janssen,
  ``{The relativistic pulsar–white dwarf binary PSR J1738+0333 – II. The
  most stringent test of scalar–tensor gravity},'' {\em Monthly Notices of
  the Royal Astronomical Society}, vol.~423, pp.~3328--3343, 07 2012.

\bibitem{PhysRevD.66.023525}
A.~Riazuelo and J.-P. Uzan, ``Cosmological observations in scalar-tensor
  quintessence,'' {\em Phys. Rev. D}, vol.~66, p.~023525, Jul 2002.

\bibitem{PhysRevD.73.083525}
A.~Coc, K.~A. Olive, J.-P. Uzan, and E.~Vangioni, ``Big bang nucleosynthesis
  constraints on scalar-tensor theories of gravity,'' {\em Phys. Rev. D},
  vol.~73, p.~083525, Apr 2006.

\bibitem{PhysRevLett.70.2217}
T.~Damour and K.~Nordtvedt, ``General relativity as a cosmological attractor of
  tensor-scalar theories,'' {\em Phys. Rev. Lett.}, vol.~70, pp.~2217--2219,
  Apr 1993.

\bibitem{PhysRevD.59.123510}
J.-P. Uzan, ``Cosmological scaling solutions of nonminimally coupled scalar
  fields,'' {\em Phys. Rev. D}, vol.~59, p.~123510, May 1999.

\bibitem{DAMOUR1994532}
T.~Damour and A.~Polyakov, ``The string dilation and a least coupling
  principle,'' {\em Nuclear Physics B}, vol.~423, no.~2, pp.~532 -- 558, 1994.

\bibitem{PhysRevLett.104.231301}
K.~Hinterbichler and J.~Khoury, ``Screening long-range forces through local
  symmetry restoration,'' {\em Phys. Rev. Lett.}, vol.~104, p.~231301, Jun
  2010.

\bibitem{khoury_chameleon_2004a}
J.~Khoury and A.~Weltman, ``Chameleon {Fields}: {Awaiting} {Surprises} for
  {Tests} of {Gravity} in {Space},'' {\em Phys. Rev. Lett.}, vol.~93,
  p.~171104, Oct. 2004.

\bibitem{khoury_chameleon_2004}
J.~Khoury and A.~Weltman, ``Chameleon cosmology,'' {\em Phys. Rev. D}, vol.~69,
  p.~044026, Feb. 2004.

\bibitem{burrage_tests_2018}
C.~Burrage and J.~Sakstein, ``Tests of chameleon gravity,'' {\em Living Rev
  Relativ}, vol.~21, p.~1, Dec. 2018.

\bibitem{BraxReview}
P.~Brax, C.~Burrage, and A.-C. Davis, ``Laboratory tests of screened modified
  gravity,'' {\em International Journal of Modern Physics D}, 06 2018.

\bibitem{burrageAtom}
C.~Burrage and E.~J. Copeland, ``Using atom interferometry to detect dark
  energy,'' {\em Contemporary Physics}, vol.~57, no.~2, pp.~164--176, 2016.

\bibitem{sabulsky_experiment_2018}
D.~Sabulsky, I.~Dutta, E.~A. Hinds, B.~Elder, C.~Burrage, and E.~J. Copeland,
  ``Experiment to detect dark energy forces using atom interferometry,'' Dec.
  2018.

\bibitem{brax_detecting_2007}
P.~Brax, C.~van~de Bruck, A.-C. Davis, D.~F. Mota, and D.~Shaw, ``Detecting
  chameleons through {Casimir} force measurements,'' {\em Phys. Rev. D},
  vol.~76, p.~124034, Dec. 2007.

\bibitem{upadhye_dark_2012}
A.~Upadhye, ``Dark energy fifth forces in torsion pendulum experiments,'' {\em
  Phys. Rev. D}, vol.~86, p.~102003, Nov. 2012.

\bibitem{chiow_multiloop_2018}
S.-w. Chiow and N.~Yu, ``Multiloop atom interferometer measurements of
  chameleon dark energy in microgravity,'' {\em Physical Review D}, vol.~97,
  Feb. 2018.

\bibitem{touboul_microscope_2017}
P.~Touboul, G.~M\'etris, M.~Rodrigues, Y.~Andr\'e, Q.~Baghi, J.~Berg\'e,
  D.~Boulanger, S.~Bremer, P.~Carle, R.~Chhun, B.~Christophe, V.~Cipolla,
  T.~Damour, P.~Danto, H.~Dittus, P.~Fayet, B.~Foulon, C.~Gageant, P.-Y.
  Guidotti, D.~Hagedorn, E.~Hardy, P.-A. Huynh, H.~Inchauspe, P.~Kayser,
  S.~Lala, C.~Lämmerzahl, V.~Lebat, P.~Leseur, F.~Liorzou, M.~List,
  F.~Löffler, I.~Panet, B.~Pouilloux, P.~Prieur, A.~Rebray, S.~Reynaud,
  B.~Rievers, A.~Robert, H.~Selig, L.~Serron, T.~Sumner, N.~Tanguy, and
  P.~Visser, ``Microscope mission: First results of a space test of the
  equivalence principle,'' {\em Phys. Rev. Lett.}, vol.~119, p.~231101, Dec.
  2017.

\bibitem{Brax:2013cfa}
P.~Brax, G.~Pignol, and D.~Roulier, ``{Probing Strongly Coupled Chameleons with
  Slow Neutrons},'' {\em Phys. Rev.}, vol.~D88, p.~083004, 2013.

\bibitem{PhysRevD.87.105013}
A.~N. Ivanov, R.~H\"ollwieser, T.~Jenke, M.~Wellenzohn, and H.~Abele,
  ``Influence of the chameleon field potential on transition frequencies of
  gravitationally bound quantum states of ultracold neutrons,'' {\em Phys. Rev.
  D}, vol.~87, p.~105013, May 2013.

\bibitem{burrage_probing_2015}
C.~Burrage, E.~J. Copeland, and E.~A. Hinds, ``Probing dark energy with atom
  interferometry,'' {\em J. Cosmol. Astropart. Phys.}, vol.~2015, no.~03,
  p.~042, 2015.

\bibitem{burrage_proposed_2016}
C.~Burrage, E.~J. Copeland, and J.~A. Stevenson, ``A proposed experimental
  search for chameleons using asymmetric parallel plates,'' {\em J. Cosmol.
  Astropart. Phys.}, vol.~2016, no.~08, p.~070, 2016.

\bibitem{ivanov_exact_2016}
A.~Ivanov, G.~Cronenberg, R.~Höllwieser, T.~Jenke, M.~Pitschmann,
  M.~Wellenzohn, and H.~Abele, ``Exact solution for chameleon field,
  self-coupled through the {Ratra}-{Peebles} potential with \$n=1\$ and
  confined between two parallel plates,'' {\em Phys. Rev. D}, vol.~94,
  p.~085005, Oct. 2016.

\bibitem{nakamura_chameleon_2018}
T.~Nakamura, T.~Ikeda, R.~Saito, and C.-M. Yoo, ``Chameleon {Field} in a
  {Spherical} {Shell} {System},'' {\em arXiv:1804.05485 [astro-ph,
  physics:gr-qc]}, Apr. 2018.
\newblock arXiv: 1804.05485.

\bibitem{kraiselburd2018}
L.~Kraiselburd, S.~J. Landau, M.~Salgado, D.~Sudarsky, and H.~Vucetich,
  ``Equivalence principle in chameleon models,'' {\em Phys. Rev. D}, vol.~97,
  p.~104044, May 2018.

\bibitem{kraiselburd2019}
L.~Kraiselburd, S.~Landau, M.~Salgado, D.~Sudarsky, and H.~Vucetich, ``Thick
  shell regime in the chameleon two-body problem,'' {\em Phys. Rev. D},
  vol.~99, p.~083516, Apr 2019.

\bibitem{hamilton_atom-interferometry_2015}
P.~Hamilton, M.~Jaffe, P.~Haslinger, Q.~Simmons, H.~Müller, and J.~Khoury,
  ``Atom-interferometry constraints on dark energy,'' {\em Science}, vol.~349,
  pp.~849--851, Aug. 2015.
\newblock arXiv: 1502.03888.

\bibitem{elder_chameleon_2016}
B.~Elder, J.~Khoury, P.~Haslinger, M.~Jaffe, H.~Müller, and P.~Hamilton,
  ``Chameleon dark energy and atom interferometry,'' {\em Phys. Rev. D},
  vol.~94, p.~044051, Aug. 2016.

\bibitem{schlogel_probing_2016}
S.~Schlögel, S.~Clesse, and A.~Füzfa, ``Probing modified gravity with
  atom-interferometry: {A} numerical approach,'' {\em Phys. Rev. D}, vol.~93,
  p.~104036, May 2016.

\bibitem{burrage_shape_2018}
C.~Burrage, E.~J. Copeland, A.~Moss, and J.~A. Stevenson, ``The shape
  dependence of chameleon screening,'' {\em J. Cosmol. Astropart. Phys.},
  vol.~2018, no.~01, p.~056, 2018.

\bibitem{ratra_cosmological_1988}
B.~Ratra and P.~J.~E. Peebles, ``Cosmological consequences of a rolling
  homogeneous scalar field,'' {\em Phys. Rev. D}, vol.~37, pp.~3406--3427, June
  1988.

\bibitem{brax_testing_2007}
P.~Brax, C.~van~de Bruck, A.-C. Davis, D.~F. Mota, and D.~Shaw, ``Testing
  chameleon theories with light propagating through a magnetic field,'' {\em
  Phys. Rev. D}, vol.~76, p.~085010, Oct. 2007.

\bibitem{sushkov_observation_2011}
A.~O. Sushkov, W.~J. Kim, D.~A.~R. Dalvit, and S.~K. Lamoreaux, ``Observation
  of the thermal {Casimir} force,'' {\em Nature Physics}, vol.~7, p.~230, Feb.
  2011.

\bibitem{lambrecht_casimir_2012}
A.~Lambrecht and S.~Reynaud, ``Casimir effect: theory and experiments,'' {\em
  International Journal of Modern Physics A}, vol.~27, p.~1260013, June 2012.

\bibitem{Llinares:2018mzl}
C.~Llinares and P.~Brax, ``{Detecting Coupled Domain Walls in Laboratory
  Experiments},'' {\em Phys. Rev. Lett.}, vol.~122, no.~9, p.~091102, 2019.

\bibitem{jaffe_testing_2017}
M.~Jaffe, P.~Haslinger, V.~Xu, P.~Hamilton, A.~Upadhye, B.~Elder, J.~Khoury,
  and H.~Müller, ``Testing sub-gravitational forces on atoms from a miniature
  in-vacuum source mass,'' {\em Nature Physics}, vol.~13, p.~938, July 2017.

\bibitem{Jain_2013}
B.~Jain, V.~Vikram, and J.~Sakstein, ``Astrophysical tests of modified gravity
  : constraints from distance indicators in the nearby universe,'' {\em The
  Astrophysical Journal}, vol.~779, p.~39, nov 2013.

\bibitem{PhysRevLett.117.101101}
A.~D. Rider, D.~C. Moore, C.~P. Blakemore, M.~Louis, M.~Lu, and G.~Gratta,
  ``Search for screened interactions associated with dark energy below the
  $100\text{ }\ensuremath{\mu}\mathrm{m}$ length scale,'' {\em Phys. Rev.
  Lett.}, vol.~117, p.~101101, Aug 2016.

\bibitem{PhysRevD.83.035020}
P.~Brax and C.~Burrage, ``Atomic precision tests and light scalar couplings,''
  {\em Phys. Rev. D}, vol.~83, p.~035020, Feb 2011.

\end{thebibliography}

\end{document}